\documentclass[floatfix,superscriptaddress,twocolumn,showpacs,aps,amsmath,amssymb]{revtex4-1}
\usepackage{graphicx}
\usepackage{amssymb}
\usepackage{soul,color}
\usepackage{bm}

\begin{document}

%\preprint{}

\title{Sample-based calibration for cryogenic broadband microwave reflectometry measurements}

\author{F. Cou\"{e}do}
 \altaffiliation[Now at ]{LPEM, CNRS, ESPCI Paris, PSL Research University, UPMC, Paris, France}
\author{L. Berg\'{e}}
\author{L. Dumoulin}
\affiliation{CSNSM-Centre de Sciences Nucl\'{e}aires et de Sciences de la Mati\`{e}re, CNRS, Univ. Paris-Sud, Universit\'{e} Paris-Saclay, 91405, Orsay, France}
\author{M. Aprili}
\affiliation{Laboratoire de Physique des Solides, CNRS, Univ. Paris-Sud, Universit\'{e} Paris-Saclay, 91405, Orsay cedex, France}
\author{C.A. Marrache-Kikuchi}
 \email{claire.marrache@csnsm.in2p3.fr}
\affiliation{CSNSM-Centre de Sciences Nucl\'{e}aires et de Sciences de la Mati\`{e}re, CNRS, Univ. Paris-Sud, Universit\'{e} Paris-Saclay, 91405, Orsay, France}
\author{J. Gabelli}
 \email{gabelli@lps.u-psud.fr}
\affiliation{Laboratoire de Physique des Solides, CNRS, Univ. Paris-Sud, Universit\'{e} Paris-Saclay, 91405, Orsay cedex, France}

\date{\today}

%§§§§§§§§§§§§§§§§§§§§§§§§§§§§§§§§§§§§§§§§§§§§§§§§§§§§§§§§§§§§§§§§§§§§§§§§§§§§§§§§§§§§§§§§§§§§§§§§§§§§§§§§§§
%Abstract
%§§§§§§§§§§§§§§§§§§§§§§§§§§§§§§§§§§§§§§§§§§§§§§§§§§§§§§§§§§§§§§§§§§§§§§§§§§§§§§§§§§§§§§§§§§§§§§§§§§§§§§§§§§
\begin{abstract}

The characteristic frequencies of a system provide important information on the phenomena that govern its physical properties. In this framework, there has recently been renewed interest in cryogenic microwave characterization for condensed matter systems since it allows to probe energy scales of the order of a few $\mu$eV. However, broadband measurements of the \textit{absolute} value of a sample response in this frequency range are extremely sensitive to its environment and require a careful calibration. In this paper, we present an \textit{in situ} calibration method for cryogenic broadband microwave reflectometry experiments that is both simple to implement and through which the effect of the sample electromagnetic environment can be minimized. The calibration references are here provided by the sample itself, at three reference temperatures where its impedance is assumed or measured, and not by external standards as is usual. We compare the frequency-dependent complex impedance (0.1--2 GHz) of an a-Nb$_{15}$Si$_{85}$ superconducting thin film obtained through this Sample-Based Calibration (SBC) and through an Open-Short-Load Standard Calibration (SC) when working at very low temperature (0.02--4 K) and show that the SBC allows us to obtain the absolute response of the sample. This method brings the calibration planes as close as possible to the sample, so that the environment electrodynamic response does not affect the measurement, provided it is temperature independent. This results in a heightened sensitivity, for a given experimental set--up.

\end{abstract}
%§§§§§§§§§§§§§§§§§§§§§§§§§§§§§§§§§§§§§§§§§§§§§§§§§§§§§§§§§§§§§§§§§§§§§§§§§§§§§§§§§§§§§§§§§§§§§§§§§§§§§§§§§§


\maketitle %\maketitle must follow title, authors, abstract and \pacs

%################################################################################################
\section{\label{sec:Intro}Introduction}
%################################################################################################

Measuring the frequency-dependent response of a system has long been a powerful mean to characterize it by determining its natural frequencies. Recently, there has been a surge in interest for microwave characterization of condensed matter systems. Indeed, microwave technology has evolved and now enables very sensitive measurements down to temperatures below 1 K. Condensed matter properties in the $\mu$eV energy range can therefore be investigated \cite{Scheffler2005, Armitage2009, Dressel2013} through this technique. One can for instance probe the intrinsic dynamics of conductors \cite{Gabelli2006, Gabelli2007} or design new measurement set-ups to determine the electrodynamic response of systems in the quantum regime, when $hf \gtrsim k_B T$ (where $h$ and $k_B$ are the Planck and Boltzmann constants respectively, $f$ is the probe frequency and $T$ the temperature), as has recently attracted much attention \cite{Sachdev1999}. However, these measurements still remain challenging due to the combination of high frequency and low temperature.

Microwave measurements can be performed using resonant cavities \cite{hafner2014surface, chen2004microwave}. The sample of interest is then inserted within the set-up and one observes how the frequency response evolves. These techniques present a high sensitivity and do not require any external calibration. There has recently been tremendous experimental progress in this field, through the development of very high quality factor superconducting resonators \cite{Kubo2011, Dassonneville2013}. However, they only work at selected frequencies corresponding to high order eigenmodes of a resonator and are mainly suitable for systems which impedance does not exhibit large variations in the considered frequency range. They also add uncontrolled parameters due to the coupling between the sample and the resonator. Broadband measurements, on the other hand, provide richer information on the system dynamical response, but are extremely sensitive to the set-up calibration. Indeed, when a microwave is sent towards a sample, the reflected or transmitted signal depends on the sample impedance $Z$, the quantity of interest, but also on the set-up itself. This constraint is easily overcome when working at room temperature where three so-called ``Standards'' can be measured before determining the sample's response. For this purpose, one usually uses an open circuit (O), a short circuit (S) and a load (L) matching the characteristic impedance of the microwave measurement setup $Z_0 =$ 50 $\Omega$. At low temperature, however, calibrating the set-up becomes more complex: temperature significantly influences physical characteristics of materials such as their impedance, dielectric constant or thermal contraction and hence changes their microwave characteristics.

Several methods have been proposed to solve this long-standing issue. Some have successively cooled down three standards at low temperature before measuring the sample\cite{Reuss2000, Kitano2008, Stutzman2000, steinberg2012broadband, scheffler2015broadband}. In the following, this will be referred to as the ``Standard Calibration'' (SC). Others considered the room-temperature calibration to be valid for two standards and used the sample at low temperature, either in the superconducting state or in a resistive state, as the third reference\cite{Booth1994, Kitano2008, Silva2016, Sarti2005}. The main drawback of both methods relies in the fact that the various cool-downs are not performed in perfectly identical conditions. In particular, they do not fully take into account the systematic errors induced by thermal gradients, which result in errors in the determination of the sample impedance. Moreover, the second procedure assumes that the sample impedance is known at one temperature. More recently, translatable microwave probes \cite{Meschede1992, Orloff2011} and electromechanical switches have been used to measure the standards and the sample at low temperature, during a single cool-down\cite{Cano2009, Yeh2013, Ranzani2013}. However, translatable probes are difficult to implement at very low temperatures ($<$ 1 K) and switches often contain magnetic parts that render them unsuitable for applications under magnetic field. In addition, these methods are sensitive to set-up imperfections: the transmission lines going to the standards and to the sample may for instance be slightly different. The sample itself is often composed of the material of interest connected to the set-up via a waveguide which response will be included in the measurement.

In this paper, we present an alternative calibration method for broadband microwave reflectometry measurements. This calibration is performed \textit{in situ} and at low temperature and requires a sample with a parameter-dependent impedance (whether temperature, magnetic field, pressure\cite{Limelette2003}, DC current, ...). The method is then based on the knowledge of the sample impedance at three different values of this parameter and allows to define the calibration plane for the measurement as close as possible to the sample. In the following, it will be referred to as the ``Sample-Based Calibration'' (SBC).

%This method only requires a single cool-down, fully takes into account the set-up temperature dependance, and can be employed under magnetic field, provided the knowledge of the sample impedance at three different temperatures. This method is particularly well suited for the study of superconducting films, amongst other topics.

In section \ref{sec:Setup}, we will detail our experimental setup. Section \ref{sec:Calibration_Principle} will outline the general principle of the calibration. We apply it to a superconducting sample, described in section \ref{sec:Sample}, which varying impedance provides a good testing ground for any calibration method. We will finally compare in section \ref{sec:New_cal} the results obtained by the SC, detailed in section \ref{sec:Std_cal}, to those provided by the SBC.
%################################################################################################

%################################################################################################
\section{\label{sec:Setup}Experimental setup}
%################################################################################################

\begin{figure}
\includegraphics[width=0.7\columnwidth]{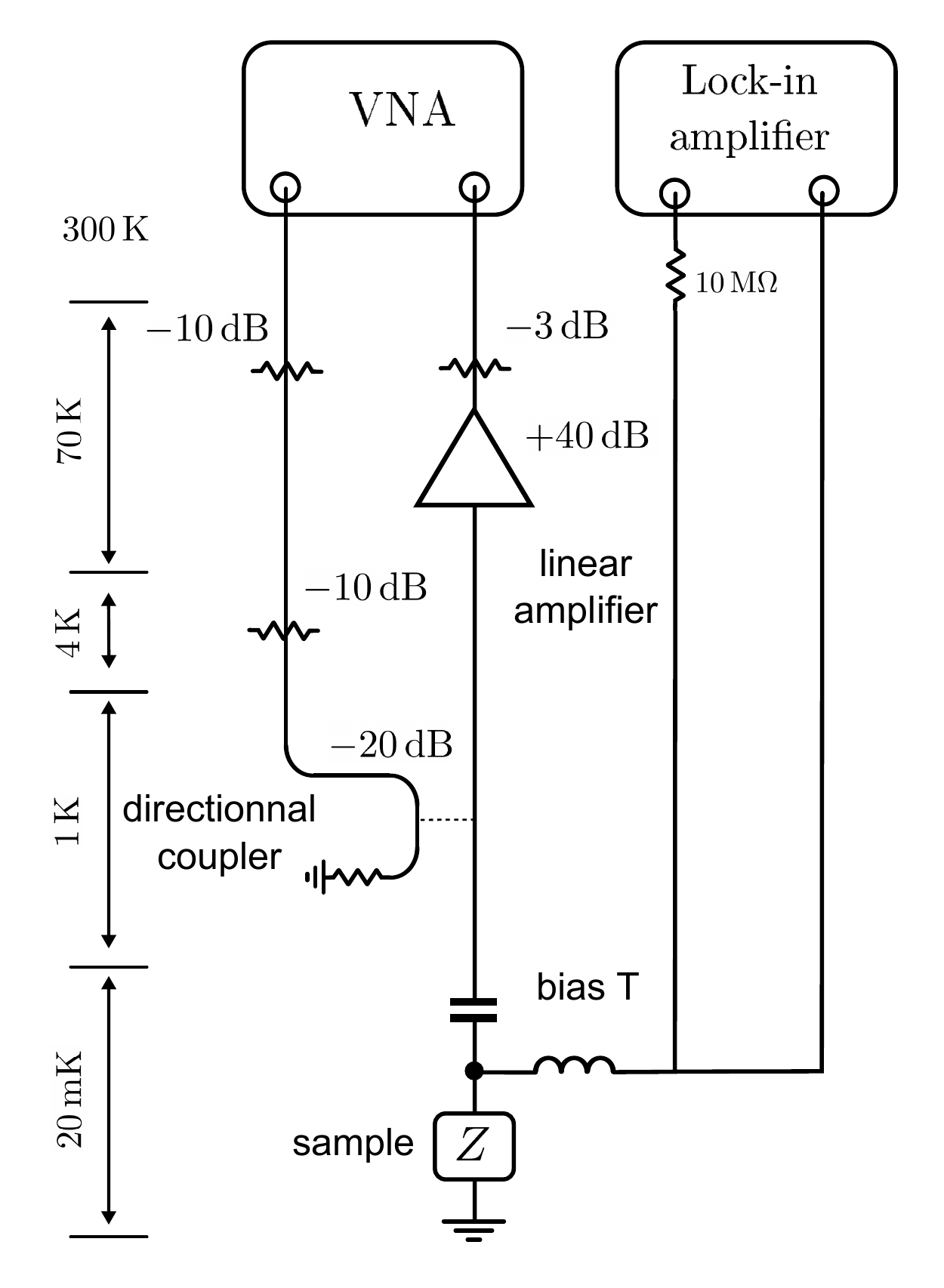}\\
  \caption{Schematic representation of the microwave reflectometry measurement setup installed in a home-made cryogen-free dilution refrigerator (until $T = 4$ K - Cryoconcept). The microwave coaxial cables are described in detail in the supplementary material section. The directional coupler is a -20 dB and 0.0001 - 2 GHz coupler from Mini-circuit company . The cryogenic amplifier is a  Miteq AFS4-00100800-22-CR-4. All the attenuators and loads comes from the XMA company and are designed to work at cryogenic temperatures.}
  \label{fig:Setup}
\end{figure}

We performed broadband microwave reflectometry measurements down to very low temperatures (T $\sim$ 20 mK) in a cryogen-free dilution refrigerator. The measurement setup is schematically shown in Fig.~\ref{fig:Setup}. Cryogenic measurements require a low excitation power to ensure thermal equilibrium between the sample and the thermal bath, resulting in a small reflected signal by the sample. Successive attenuations of the incoming signal are thus needed at the different stages of the refrigerator to minimize the thermal noise arriving on the sample, while the sample's reflection has to be subsequently amplified. These two conditions are met by using a directional coupler at $T = 1$ K, which allows us to decouple the excitation and detection microwave lines. The microwave power is delivered by a Rohde $\&$ Schwarz ZVB Vector Network Analyzer (VNA), which also measures the reflected signal $\Gamma^m$. The bandwidth of this setup is set at low frequency, $f_{min}$ = 100 MHz, by the cryogenic HEMT amplifier (Miteq AFS4-00100800-22-CR-4, noise temperature $T_N\simeq$ 70 K \cite{Gabelli2008}), while the upper frequency limit, $f_{max}$= 2 GHz, is fixed by the directional coupler (Mini-circuit ZFDC-20-5+). \\

%As mentioned earlier, the set-up characteristics are temperature-dependent. Most notably, the signal attenuation is reduced when the temperature is lowered due to the decrease of the resistiviy of the materials in the coaxial cables.

A low-frequency ($f_{lock-in} \sim$ 77 Hz) bias is simultaneously applied to the sample by a lock-in amplifier through a bias tee to measure the sample low frequency resistance. Measurements are performed in the linear regime where both the microwave power ($\sim 10$ fW on the sample) and the lock-in amplifier current ($\sim$ 1 $\mu$A) are sufficiently low to prevent heating the sample.

%################################################################################################

%################################################################################################
\section{\label{sec:Calibration_Principle}Calibration Principle}
%################################################################################################

In broadband reflectometry, the attenuation of microwave lines is temperature-dependent and the set-up therefore requires a calibration to relate the measured reflected signal to the signal actually reflected by the sample only \cite{Kitano2008,Booth1994}.
%the measured reflection coefficient $\Gamma$ is a function of the reflection coefficient of the sample $\Gamma_{sample}$, but also of the entire measurement apparatus.
Indeed, the cables as well as the various interfaces between connectors modify the amplitude and the phase of the propagating signal. The point of any calibration method is to characterize these extrinsic effects.

\begin{figure}
\includegraphics[width=0.7\columnwidth]{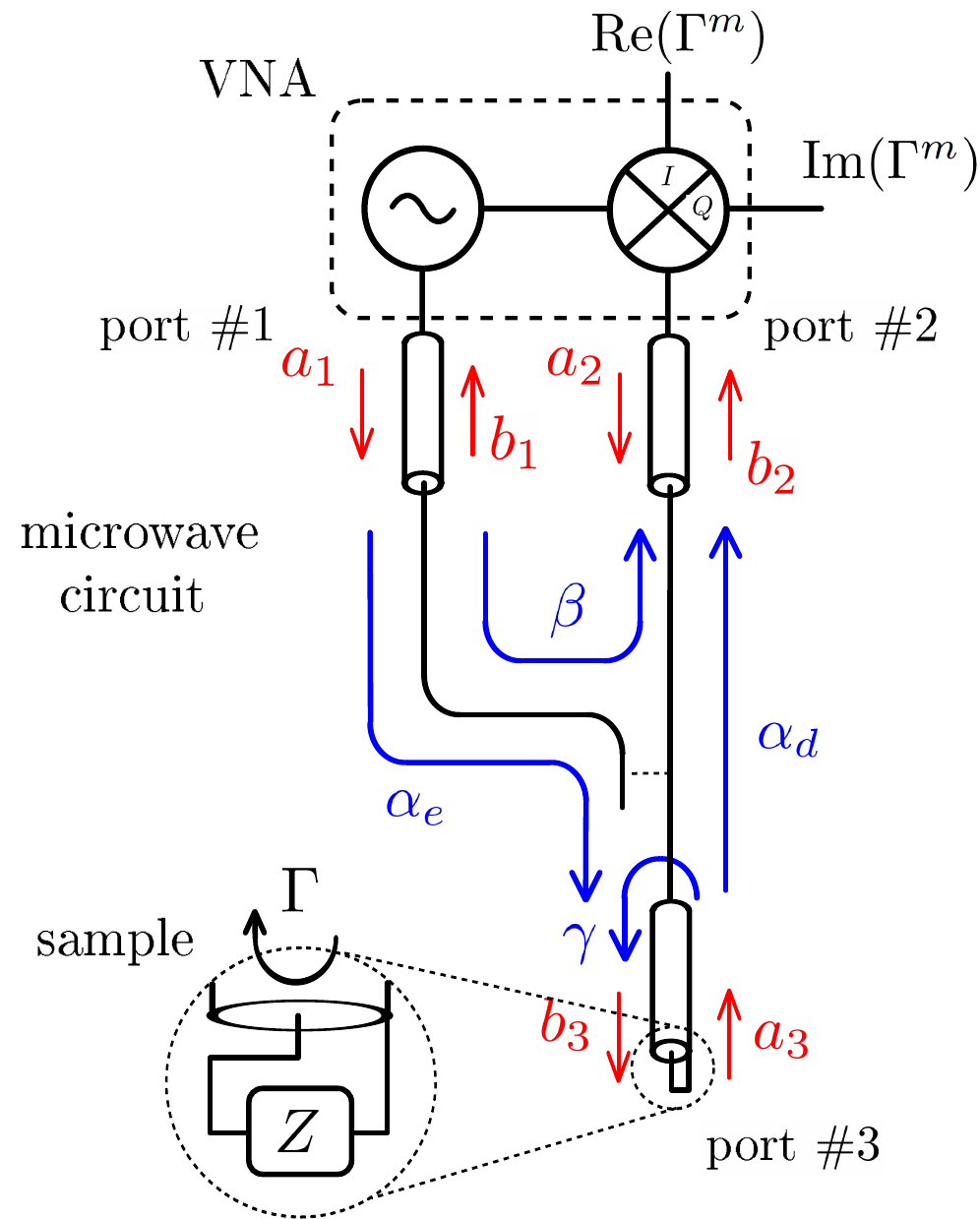}\\
  \caption{(color online) Schematic representation of the reflectometry measurement apparatus. The microwave source is at port \#1, and the reflected signal is detected at port \#2. The sample is represented by a complex impedance $Z$ at port \#3. }
  \label{fig:Error_coeff}
\end{figure}

The reflectometry set-up depicted Fig.~\ref{fig:Setup} can be modeled by a three-ports system, as shown in Fig.~\ref{fig:Error_coeff}. Ports \#1 and \#2 correspond to the source and the detector of the VNA, respectively, while the sample is connected to port \#3, generally through a sample holder. At each port \textit{i}, the incoming and outgoing waves amplitudes are labeled $a_i$ and $b_i$, respectively. These complex waves are related through four calibration coefficients: $\alpha_e$ and $\alpha_d$ describe the transmission of the excitation and detection lines respectively; $\beta$ represents the parasitic coupling between the excitation and detection lines due to the imperfections of the directional coupler insulation; whereas $\gamma$ corresponds to part of the signal being reflected back to port \#3, originating from the impedance mismatch of the microwave line connected to the sample.

The reflection coefficient, $\Gamma^m = \frac{b_2}{a_1}$, measured at the VNA, is then related to the reflection at port \#3, $\Gamma=\frac{b_3}{a_3}$ by :
\begin{equation}
    \Gamma^m = \beta+\frac{\alpha\Gamma}{1-\gamma \Gamma}
    \label{eq:Reflection_coeff}
\end{equation}
\noindent Note that $\alpha=\alpha_e \alpha_d$ reflects the attenuation of the entire line. Moreover, if $Z_{\#3}$ is the complex impedance of the elements between port \#3 and the ground, one has:
\begin{equation}
   \Gamma=\frac{Z_{\#3}-Z_0}{Z_{\#3}+Z_0}
   \label{eq:Gamma_sample}
\end{equation}

From equation (\ref{eq:Reflection_coeff}), it is clear that three independent complex coefficients ($\alpha$, $\beta$, $\gamma$) are needed in order to retrieve $\Gamma$ from the measurement of $\Gamma^m$. Let us emphasize that these coefficients depend not only on the length of the cables or on their materials, but also on the quality of the connection of the various connectors, and, as highlighted above, on the temperature. For each experimental configuration, one therefore needs a calibration comprising three independent measurements of known references in order to determine $\Gamma$, taken from the calibration plane. Equation (\ref{eq:Gamma_sample}) then enables to determine $Z_{\#3}$.

Upon performing a Standard Calibration using references that are successively cooled down, one therefore assumes that $\alpha$, $\beta$ and $\gamma$ do not vary from one cool-down to the next. For this assumption to be realistic, great experimental care has to be taken to reproduce as exactly as possible the different temperature gradients in the set-up, the mechanical stress on the different components or the various microwave connections. Mastering the identical reproduction of these experimental conditions is all the more difficult that the operating temperature is low. In the SBC we propose, we make the hypothesis that the set-up is temperature independent from an electrodynamic point of view. In our case, this is verified at least for $T<4$ K. Moreover, the reference will be the sample itself at three different temperatures where its response can reasonably be assumed to be known. If those two pre-requisites are met, the calibration can be performed in a single cool-down and with experimental conditions which similarity is only limited by the drift of the measurement set-up in time.

Equations (\ref{eq:Reflection_coeff}) and (\ref{eq:Gamma_sample}) show that the calibration allows to take into account the entirety of the sample's environment, except for what is inserted in between port \#3 and the ground. In usual experiments, this interval is not only occupied by the sample of interest itself, but also by connectors, electrical leads, a substrate, links with the ground, etc. The electrodynamic response of these various components are therefore, by construction, included in $\Gamma$ and indistinguishable from the sample's. The way to circumvent this is to propose an electrodynamic model or to vary an experimental parameter, such as the temperature or the magnetic field, to help disentangle the different elements. In our case, when the SC is performed using Open-Short-Load standards, port \#3 corresponds to the plane where the SMA connects with the sample holder and where the standards and the sample are successively inserted. By contrast, in the SBC, the only element that varies between the references and the sample measurements is the sample itself. This means that all surrounding elements are integrated in the calibration process. In other words, port \#3 corresponds to the plane at the immediate vicinity of the sample.

In the following, we will compare results obtained on a superconducting sample using a SC and a SBC. The SBC references will consist in two pure resistances (sample in the normal state) and a pure inductance (sample at low temperature). Before proceeding, we would like to stress that our experimental set-up has not been optimized for the SC. State-of-the-art SC at mK temperatures using electromechanical switches has been reported in \cite{Ranzani2013}. The point of the comparison between the two calibration methods is to show that, whenever possible, for a given experimental set-up, using the SBC may be easier to implement. Furthermore, the SBC has a built-in enhanced resolution due to the fact that it effectively removes the effect of the sample's environment as opposed to assuming it to be negligible.

%In the following, we will compare results obtained using two different sets of references. For the SC, we will consider the Open-Short-Load standards. In the SBC we propose, the references will be the sample itself at three different temperatures where its response can reasonably be assumed to be known. We chose to apply these calibration methods onto a superconducting sample, so that the SBC references will consists in two pure resistances (sample in the normal state) and a pure inductance (sample at low temperature).

%. In addition to the SC, such a system will allow us to perform a SBC by considering pure resistances (sample in the normal state) and a pure inductance (at low temperatures).
%################################################################################################

%################################################################################################
\section{\label{sec:Sample}Sample description}
\begin{figure}
\includegraphics[width=0.9\columnwidth]{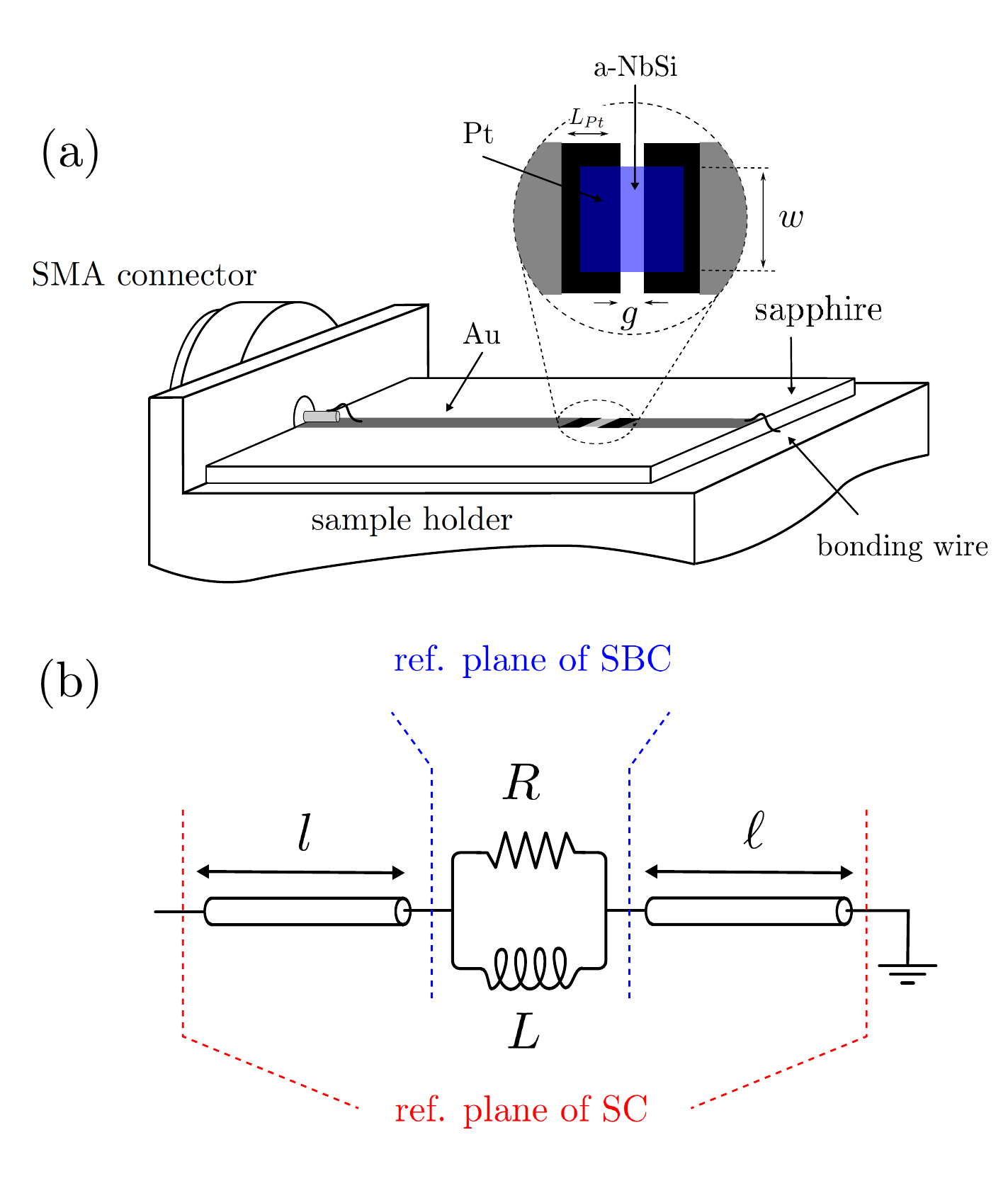}\\
  \caption{(color online) (a) Schematic representation of the sample geometry: the NbSi film is placed within a micro-strip line. (b) Electromagnetic model of the superconducting sample shown in (a): the thin film of interest lies in between two transmission lines. The dotted lines correspond to the calibration planes for the SC (red) and the SBC (blue). See text.}
  \label{fig:Sample}
\end{figure}

%################################################################################################
A schematic representation of the sample is shown in Fig.~\ref{fig:Sample}(a). It is placed within a transmission line consisting in a 400 $\mu$m-wide gold microstrip (of thickness 200 nm) deposited at the surface of a 500 $\mu$m-thick sapphire substrate. We chose sapphire in order to have a good thermalization of the sample at low temperatures. Moreover, ceramics have a surface roughness that prohibits their use for thin films. At the back of the substrate, a gold plane was deposited to ensure a good electrical contact with the sample holder ground. The width of the transmission line has been chosen so that the overall line impedance is close to $Z_0$. Let us emphasize that the microstrip geometry is very flexible to meet this impedance matching condition for any given sample resistivity, provided the lumped element approximation is valid. One end of the transmission line is connected to the measurement setup through a SMA-type connector, which pin is directly wire-bonded to the microstrip via multiple 25 $\mu$m-wide gold wires. The other end is wire-bonded to the ground.
%The transmission line itself is composed of three parts.  \sout{The incoming microwave signal is generated by the source of a network analyzer\cite{Note_VNA} and is sent through a coaxial cable terminated by a SMA-type connector} into the first segment, a 200 nm-thick gold line of length $L_1=3$ mm. The connector pin is wire-bonded to the gold line via multiple 25 $\mu$m-wide gold wires. $L_1$ has been chosen to provide sufficient space for confortable wire-bonding. The central part consists in the sample itself. Since we are measuring the sample reflection, the other end of the sample is grounded through a gold line of length $L_2=1.5$ mm. There again, $L_2$ is sufficiently large to enable numerous wire-bonding to the sample holder ground.

The sample here consists in an amorphous Nb$_{15}$Si$_{85}$ (a-NbSi) superconducting thin film of thickness $d$=12.5 nm\footnote{The a-NbSi film has been prepared at room temperature and under ultrahigh vacuum (typically a few 10$^{-8}$ mbar) by electron beam co-deposition of Nb and Si, at a rate of the order of 1 \AA.s$^{-1}$ as described in \cite{Couedo2016}.}. In order to prevent any diffusion of gold into the a-NbSi film, a 25 nm-thick Pt buffer layer was inserted between the gold line and the sample as shown in the blow-up Fig.~\ref{fig:Sample}.a: on both sides of the samples the length of the Pt line is $L_{Pt}=195$ $\mu$m. Indeed, platinum does not diffuse in a-NbSi and has the advantage of being a good conductor, non-magnetic -- which is important when dealing with superconducting samples -- and non-oxidizable to ensure a good ohmic contact with the thin film. The geometry of the a-NbSi film was tuned so that its normal state resistance ($R_n\simeq 25 \, \Omega$) is close to $Z_0$: its width was $w=385$~$\mu$m and its effective length $g=10$~$\mu$m, which corresponds to the length of the NbSi film that becomes superconducting at low temperature. Indeed, we have checked in a separate experiment that, in regions where Pt and a-NbSi were superimposed, superconductivity was suppressed by inverse proximity effect, allowing for a clear definition of the calibration planes (see Fig.\ref{fig:Sample}(b)). Both the microstrip line and the sample were fabricated using standard photo-lithography and etching techniques.

Before any low temperature reflectometry measurements, we have characterized the room temperature frequency response of the sample by using a SC at port \#3. As the incoming microwave arrives on the sample, the main impedance mismatch encountered is at the interface between the gold microstrip line and the sample. The reflection coefficient $\Gamma$ is then associated with a Fabry-Perot type resonator made out of the sample and of the second part of the gold microstrip line of length $\ell \simeq 2.0 \pm 0.25 \, \mathrm{mm}$. It includes the $1.5 \,\mathrm{mm}$-long line between the sample and the grounded plane and the length of the bonding wires connecting this plane to the ground. The phase $\varphi$ associated with the propagation of the electromagnetic wave of frequency $\omega$ between the a-NbSi film and the ground is then given by $\varphi=\frac{\omega \ell}{c}$ with $c \simeq 1.1 \times 10^8 \, \mathrm{m.s}^{-1}$, the velocity of the electromagnetic wave in the microstrip. In the same way, a global phase $\phi=\frac{\omega l}{c}$ is associated with the propagation along the SMA connector and the first part of the gold microstrip line, where $l$ stands for an effective propagation length that includes both the real microstrip line length and the effects of the SMA connector. $Z_{\#3}$ then corresponds to the total impedance $Z_{FP}$ of the Fabry-Perot resonator, which is then related to the reflection coefficient according to:
\begin{equation}
    \Gamma=\frac{Z_{FP}-Z_0}{Z_{FP}+Z_0}=\left(\frac{\left(Z-Z_0\right) \cos \varphi-iZ_0 \sin \varphi}{\left(Z+Z_0\right) \cos \varphi+iZ_0 \sin \varphi}\right)\text{e}^{i\phi}
\label{eq:Z300K}
\end{equation}
\noindent where $Z$ is the impedance of the a-NbSi film.

\begin{figure}
\includegraphics[width=0.85\columnwidth]{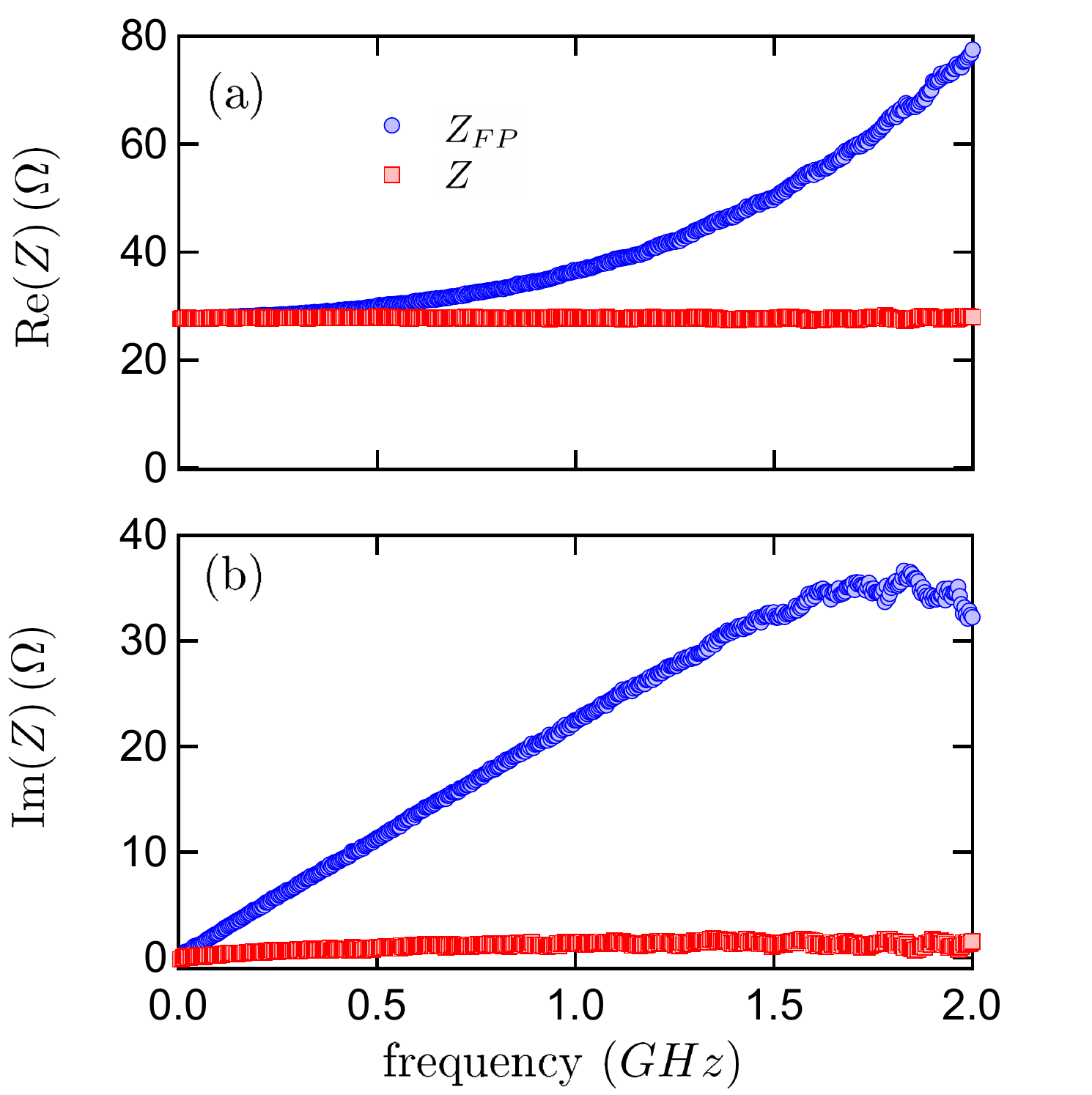}\\
 \caption{(color online) Real (a) and imaginary (b) parts of the complex impedances $Z_{FP}$ (blue symbols) and $Z$ (red symbols) at room temperature. $Z$ corresponds to the a-NbSi thin film impedance, obtained via equation (\ref{eq:Z300K}) with $l=25 \, \mathrm{mm}$ and $\ell=2.0 \, \mathrm{mm}$.}
  \label{fig:RT_charac}
\end{figure}

This electromagnetic model for the sample can be tested at room temperature. Fig.~\ref{fig:RT_charac} shows the frequency dependance of both real and imaginary parts of $Z_{FP}$. At low frequency, they are in good agreement with the sample impedance as measured by standard lock-in techniques: $\left.\text{Re}(Z_{FP})\right|_{\omega\rightarrow0}=\text{Re}(Z)=27.7 \,\Omega$ and $\left.\text{Im}(Z_{FP})\right|_{\omega\rightarrow0}=\text{Im}(Z)=0 \, \Omega$. These values are consistent with what is expected for a metallic thin film, which geometry is such that we can neglect both the geometric inductance ($\sim 20$ pH) and the capacitance to the ground ($\sim 1$ fF). The sample is also sufficiently short to neglect any propagation effect for frequencies below a few hundreds MHz. At room temperature and low frequency, the a-NbSi thin film therefore behaves itself as a pure resistor. At higher frequencies, however, the effect of propagation in the Fabry-Perot resonator can be seen and the total impedance is no longer purely real. We can retrieve $Z$ from $Z_{FP}$, using equation~(\ref{eq:Z300K}). $\ell=2.0 \pm 0.25 \, \mathrm{mm}$ has been set by the geometry of the sample. $l$ was the only fitting parameter and has been adjusted so that $\text{Re}(Z)$ and $\text{Im}(Z)$ are frequency-independent on the whole frequency range. As can be seen, there is an excellent agreement between model and experiment for $l=25 \pm 0.5 \, \mathrm{mm}$ which is a realistic value for our setup. Moreover, it should be stressed that no additional capacitance $C$ was needed in the model to reproduce the data, in agreement with the result of a microwave simulation giving an upper limit of $C < 50\, \mathrm{fF}$ for the gap of $g=10\, \mu\mathrm{m}$ in the microstrip line. This also rules out any parasitic effect coming from sample fabrication (remaining resist or contact resistances for eg.). At room temperature, the electromagnetic response of the ensemble $\{$microstrip + sample$\}$ is therefore very well described by the sample resistance and Fabry-Perot type effects, as modeled by equation (\ref{eq:Z300K}) and it enabled us to extract the value of the sample impedance at room temperature. In the following, $l=25 \, \mathrm{mm}$ and $\ell=2.0 \, \mathrm{mm}$ will be considered as fixed parameters.

%We have therefore considered that the electromagnetic model presented figure \ref{fig:RT_charac}.a is realistic and could therefore exclude any parasitic effects. These could for instance arise from sample fabrication such as additional capacitances due to remaining resist. %Geometrical capacitances are negligible, in accordance with the numerical simulations of the sample.

%The electromagnetic model presented figure \ref{fig:RT_charac} is therefore realistic and has been validated at room temperature. From now on, we will focus on the low temperature situation.

In the rest of this paper, we will focus on the temperature and frequency dependences of the impedance $Z(T,\omega)$ in the low frequency limit $\hbar \omega\ll \Delta_0$, where $\Delta_0=1.76k_BT_c$ is the superconducting energy gap. Below the critical temperature $T_c=0.23\, \mathrm{K}$, the sample is superconducting and, within the two-fluid model approximation, behaves like a parallel $RL$ circuit which parameters we will now evaluate. The a-NbSi sample is characterized by its electronic mean free path $l_e$, its coherence length $\xi$ and its London penetration depth $\lambda_0$ :
\begin{eqnarray*}
l_e &=& \frac{3D}{v_F}\sim 4 \, \text{\AA} \\
\xi&=&\sqrt{\xi_0l_e} = \sqrt{\frac{\hbar v_Fl_e}{\pi \Delta_0}} \sim 35 \, \mathrm{nm}\\
\lambda_0 &=& \sqrt{\frac{m}{\mu_0 n e^2}} \sim 100\, \mathrm{nm}
\end{eqnarray*}
\noindent where $D \sim 6\times 10^{-5}\, \mathrm{m^2.s}^{-1}$ is the diffusion constant\cite{Aubin2006}, $v_F \sim 0.5 \times 10^6 \, \mathrm{m.s}^{-1}$ is the Fermi velocity and $n \sim 5 \times 10^{27} \, \mathrm{m}^{-3}$ the electron density. The Cooper pairs kinetic inductance at $T=0 \, \mathrm{K}$ is then given by BCS theory in the dirty limit ($l_e \ll \xi, \lambda_0$):
\begin{equation}
L_K(T=0)=\frac{\hbar R_{n}}{\pi\Delta_0} \simeq 140 \, \mathrm{pH}
\label{eq:Ls(T=0)}
\end{equation}
where $R_n = R(4\, \mathrm{K}) \simeq$ 26.5 $\Omega$ is the normal state resistance of the sample. Since $d\ll \lambda_0$ and since the dissipative part of the conductance is only due to unpaired electrons \footnote{The dissipation given by the Mattis-Bardeen theory, related to thermally excited quasiparticles, will be discussed in Fig.~\ref{fig:Z_SBC}(a).}, the electromagnetic field penetrates all the sample and the parallel $RL$ circuit in the low frequency limit is characterized by \cite{Tinkham}:
\begin{eqnarray}
R(T) &=& R_n \exp \left(\Delta(T)/k_BT\right) \label{eq:RL1}\\
L(T) &=& \frac{\hbar R_n/\pi\Delta(T)}{\tanh \left( \Delta(T)/2k_BT \right)} \label{eq:RL2}
\end{eqnarray}
\noindent where $\Delta(T)$ stands for the temperature dependence of the superconducting gap. In the following, although $\Delta(T)$ should, strictly speaking, be given by the self-consistent BCS gap equation, we will use the interpolation formula $\Delta(T)=\Delta_0 \tanh \left(1.74 \sqrt{T_c/T-1}\right)$ valid for $T$ close to $T_c$. The overall complex impedance of the sample can thereby be written as:
\begin{equation}
Z = \frac{RL^2\omega^2}{R^2+\left(L\omega\right)^2}+j\frac{R^2L\omega}{R^2+\left(L\omega\right)^2}
\label{eq:Z_BCS}
\end{equation}
\noindent We will now examine how the measured sample impedance compares with equation (\ref{eq:Z_BCS}) for both the SC and the SBC. We should stress that the expression of $R(T)$ given by equation (\ref{eq:RL1}) is nothing more than the Drude resistance and does not account for the dissipation related to thermally excited quasiparticles at finite frequencies. A more detailed study would use Mattis-Bardeen theory to describe the conductance of the superconducting film \cite{Mattis1958}, but we will see in section \ref{subsec:Results} that despite its simplicity, the two-fluid model captures the major part of the system's physics.

%################################################################################################

%################################################################################################
\section{\label{sec:Std_cal}Standard calibration procedure}
%################################################################################################

\begin{figure}
\includegraphics[width=0.85\columnwidth]{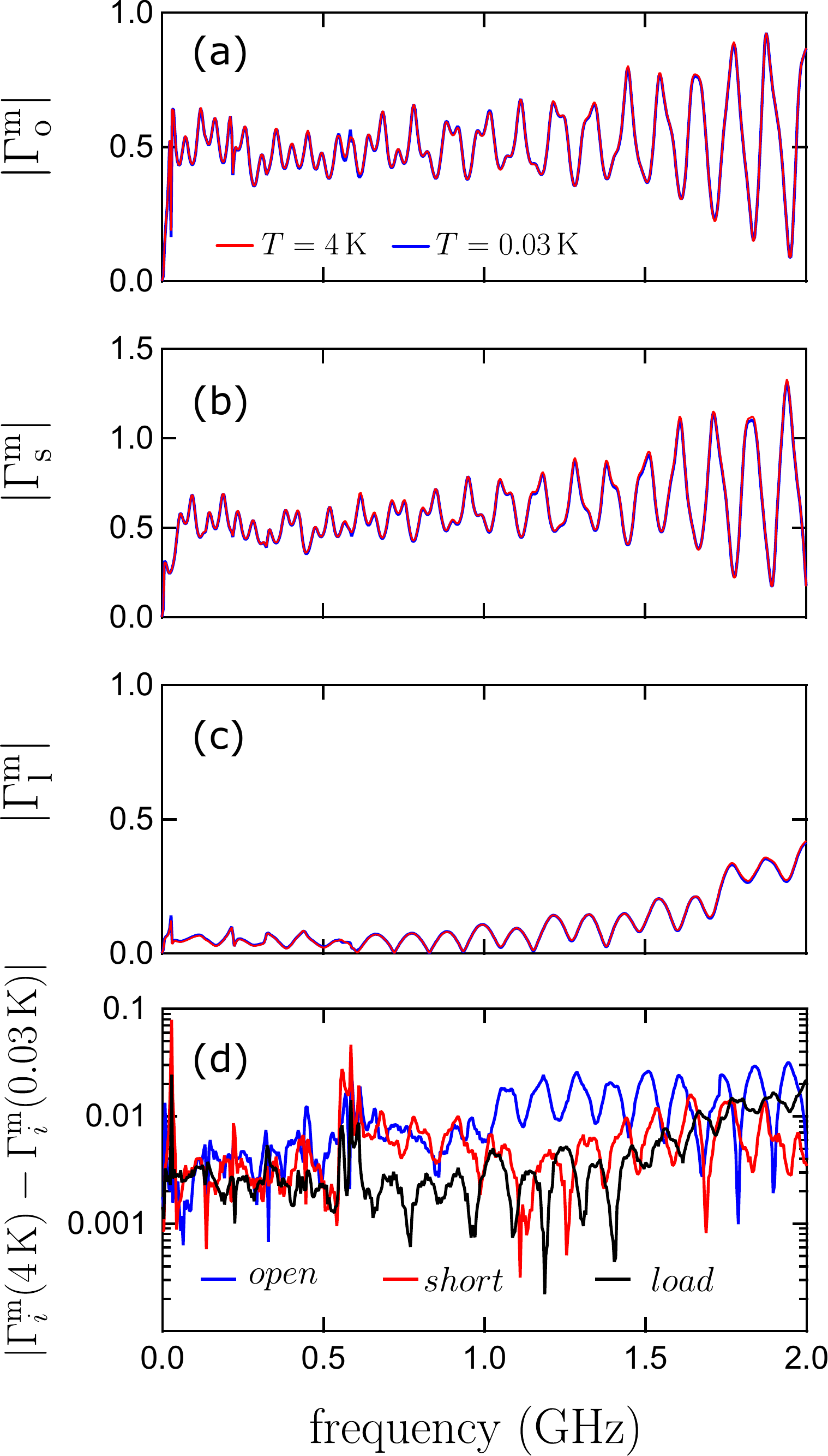}\\
  \caption{(color online) Frequency dependence of the magnitude of the reflexion coefficient for the open (a) short (b) and load (c) standards used for the SC, at $T = 4$ K and $T = 0.03$ K. Panel (d) shows the change in reflexion between these two temperatures for each of these standards ($ i=\{o,s,l\}$).}
  \label{fig:standard_T}
\end{figure}

\begin{figure*}
\includegraphics[width=1.35\columnwidth]{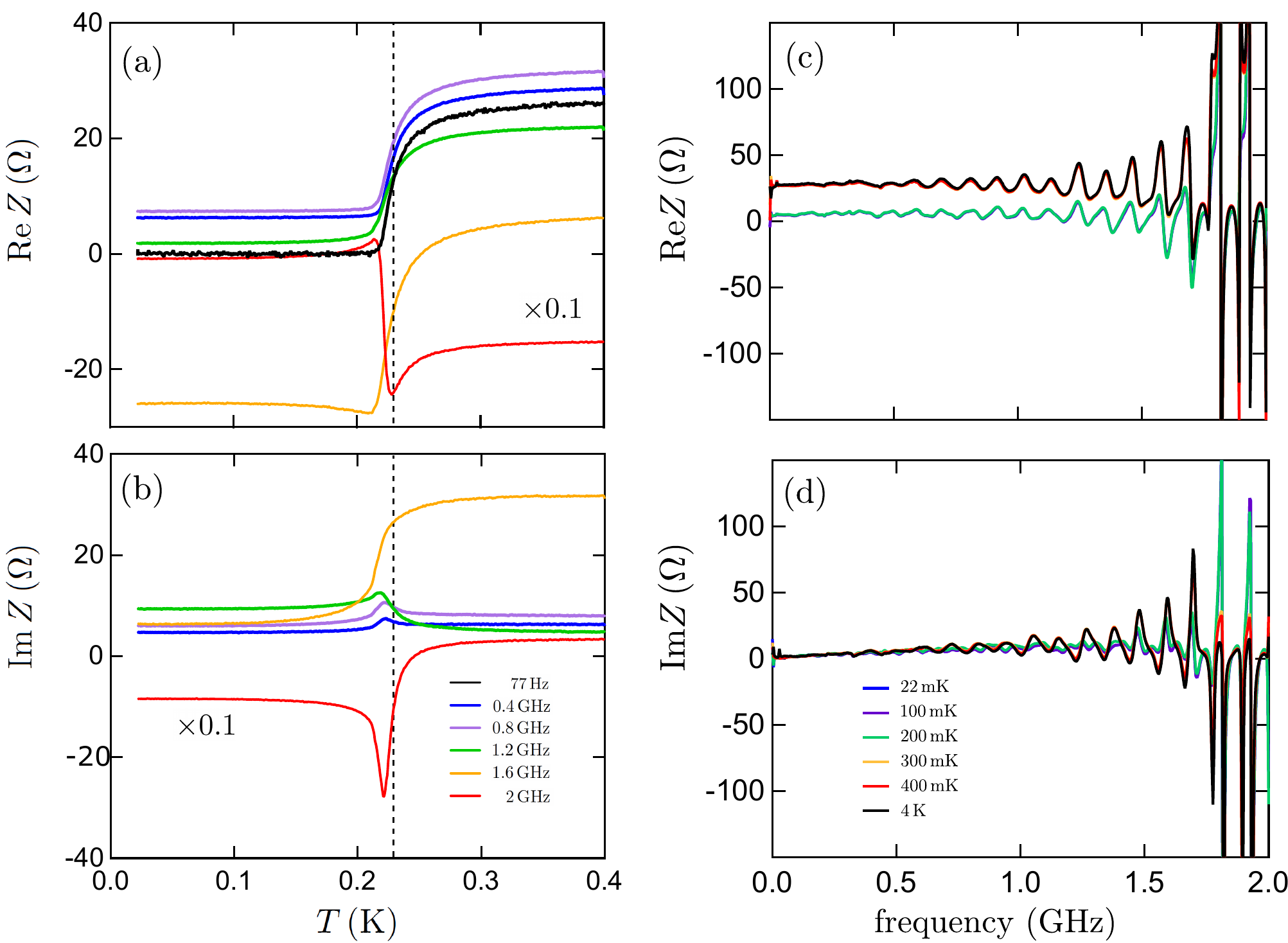}\\
  \caption{(color online) (a) Real and (b) imaginary parts of the complex impedance as a function of temperature for selected frequencies, as obtained from a standard calibration (see text). Re($Z$) is also compared to the lock-in measurement. Measurements at $2 \, \mathrm{GHz}$ has been multiplied  by a factor  $0.1$  for clarity. Real (c) and imaginary (d) parts of the complex impedance as a function of frequency for selected temperatures.}
  \label{fig:Z_SC}
\end{figure*}

At first, let us consider the SC in which three references, an open, a short and a load (see Supplementary Material for more details on these standards), were successively connected to port \#3 and cooled down at $4 \,\mathrm{K}$ in order to determine the error coefficients, as described in \cite{Kitano2008,Reuss2000}. %(open $Z=+\infty$, short $Z=0$ $\Omega$, load $Z=50$ $\Omega$).
Once again, let us stress that our experimental set-up is not optimized for this measurement and that we have used this method solely to be able to compare it to our SBC on the same set-up.
%We developed this method in order to compare it with our SBC. However, we would like to stress that our experimental set-up is not optimized for this measurement. State-of-the art SC at mK temperatures using electromechanical switches has been reported in .
Figure~\ref{fig:standard_T} shows the frequency dependence of the reflexion coefficients for each of the standards at low temperature. As can be seen, the set-up response scarcely varies, within experimental uncertainty ($ |\delta \Gamma^m| \lesssim 5\times 10^{-3}$) in the 30 mK - 4 K. The sample was then measured during a fourth cool-down. In the SC procedure, the $\alpha$, $\beta$ and $\gamma$ coefficients are directly given by:
\begin{eqnarray*}
    \alpha&=&2\frac{\left(\Gamma^m_{l}-\Gamma^m_{s}\right)\left(\Gamma^m_{o}-\Gamma^m_{l}\right)}{\Gamma^m_{o}-\Gamma^m_{s}}\\
    \beta&=&\Gamma^m_{l}\\
    \gamma&=&1+2\frac{\Gamma^m_{s}-\Gamma^m_{l}}{\Gamma^m_{o}-\Gamma^m_{s}}
\end{eqnarray*}
\noindent where $\Gamma^m_o$, $\Gamma^m_s$ and $\Gamma^m_l$ correspond to the measured reflection coefficient of the Open, Short and Load standards respectively. Using equations (\ref{eq:Reflection_coeff}), (\ref{eq:Gamma_sample}) and (\ref{eq:Z300K}), we retrieved the temperature and frequency dependences of the real and imaginary parts of the sample impedance, Re($Z$) and Im($Z$) (Fig.~\ref{fig:Z_SC}).

\begin{figure*}
\includegraphics[width=1.3\columnwidth]{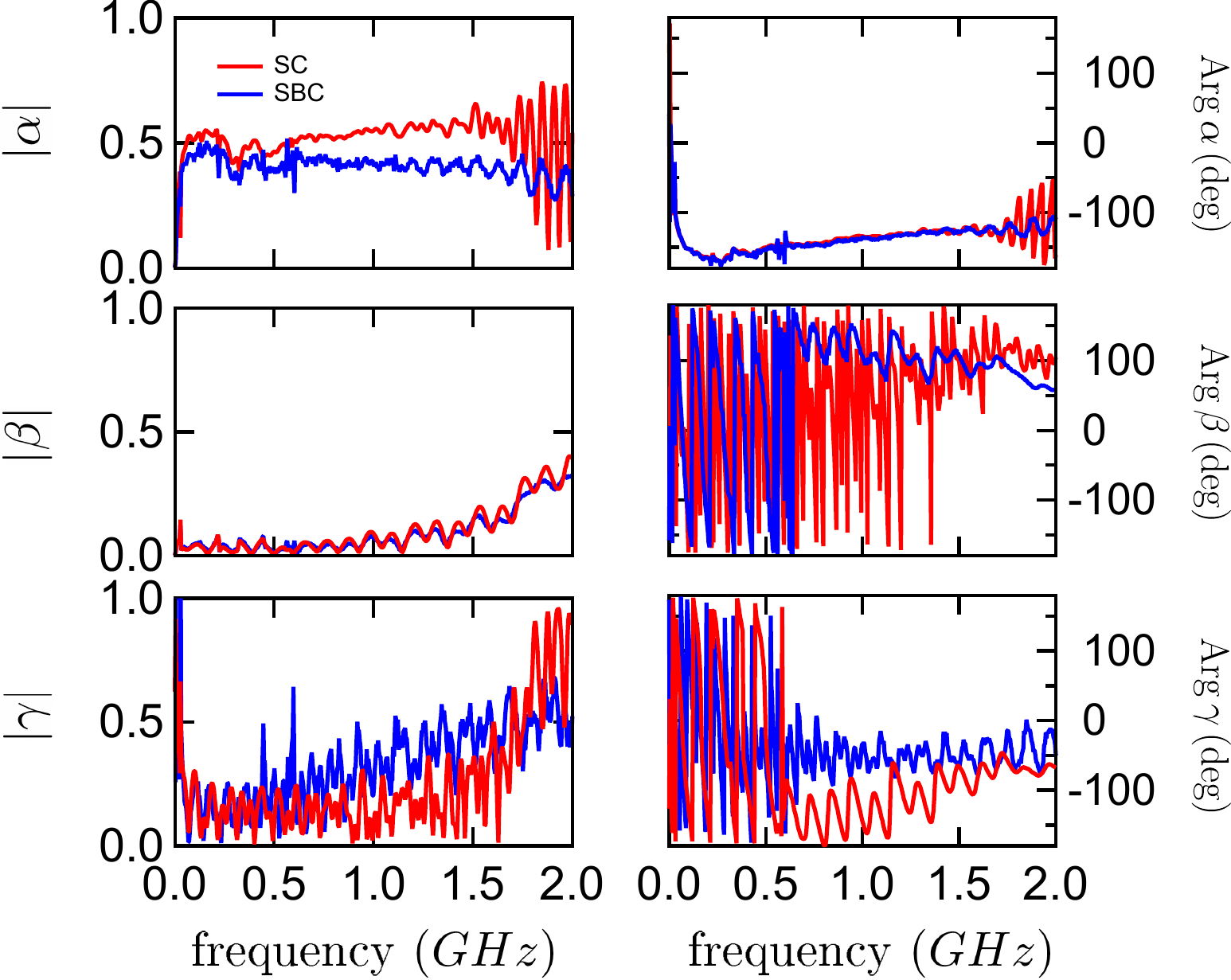}\\
  \caption{(color online) Calibration coefficients $\alpha$, $\beta$ and $\gamma$ for the SC (red) and SBC (blue) procedures. Left and right panels show the magnitude and the phase of these coefficients, respectively. For the SC calibration, the error coefficients have been determined from measurements of the OSL standards at $T = 4$ K.}
  \label{fig:coeff_cal}
\end{figure*}

As can be seen, Re($Z$) follows qualitatively the expected temperature dependence at low frequency ($f < 1\, GHz$)\cite{Mattis1958, Driessen2012}. In particular, $T_c$ is clearly identifiable through the drop in Re($Z$). Concomitantly, Im($Z$) displays a maximum near $T_c$, as expected from equation (\ref{eq:Z_BCS}). However the low temperature values of both components exhibit a frequency-dependent offset that bears no physical ground. This is partly due to irreproducibility inherent to any cool-down. Thermal contractions, such as those in microwave cables or the contraction of the insulating dielectric in SMA-type connectors, may differ. Even in a cryogen-free dilution refrigerator, thermal gradients may be ill-controlled. These effects in turn give rise to cool-down-dependent impedance mismatches that are not taken into account by this calibration method. Moreover, the OSL standards may also have a different response at low temperature \footnote{The evolution in the standards frequency response at low temperature was not corrected, although we have checked that the impedance of the load was still 50 $\Omega$ at low temperature.}. Finally, we have used parameters for the Fabry-Perot resonator ($l$, $\ell$) that have been determined at room temperature and these may be slightly temperature-dependent. These effects are illustrated Fig.~ \ref{fig:coeff_cal} where Fabry-Perot-type oscillations in the magnitude of $\alpha$ and $\gamma$ are flagrant at high frequency. As a result, the phase reference is ill-defined and the real and imaginary parts of $Z$ are mixed up (Fig.~\ref{fig:Z_SC}(c), (d) and $Z(f=2\, \mathrm{GHz})$ in Fig.~\ref{fig:Z_SC}(a), (b)).

Although our SC satisfactorily describes the \textit{qualitative} features of the superconductor response at low frequency and might therefore be adequate to determine \textit{relative variations} in $Z$, it is not appropriate to finely measure the  \textit{absolute} value of the sample impedance. As we will show in the following section, it is necessary to use a more precise calibration procedure to probe the frequency dependence of the dynamical response for a superconducting film.

%\begin{figure}[t!]
%\includegraphics[width=0.9\columnwidth]{fig_gamma_T_B}\\
%  \caption{(color online) Magnitude of the reflection changes for the Open ($o$), Short ($s$) and Load ($l$) standards when increasing the temperature from $T = 33$ mK to $4$ K (red curves) and when applying a magnetic field $B$ - perpendicular to the sample holder - from $B = 0$ T to $1$ T, at a fixed temperature $T = 33$ mK (blue curves).}
%  \label{fig:setup_variationT_B}
%\end{figure}

%################################################################################################

%################################################################################################
\section{\label{sec:New_cal}Sample-based calibration procedure}
%################################################################################################
%-----------------------------------------------------------------------------------

\subsection{\label{subsec:Gal_pple}General principle}
\begin{figure}
\includegraphics[width=0.85\columnwidth]{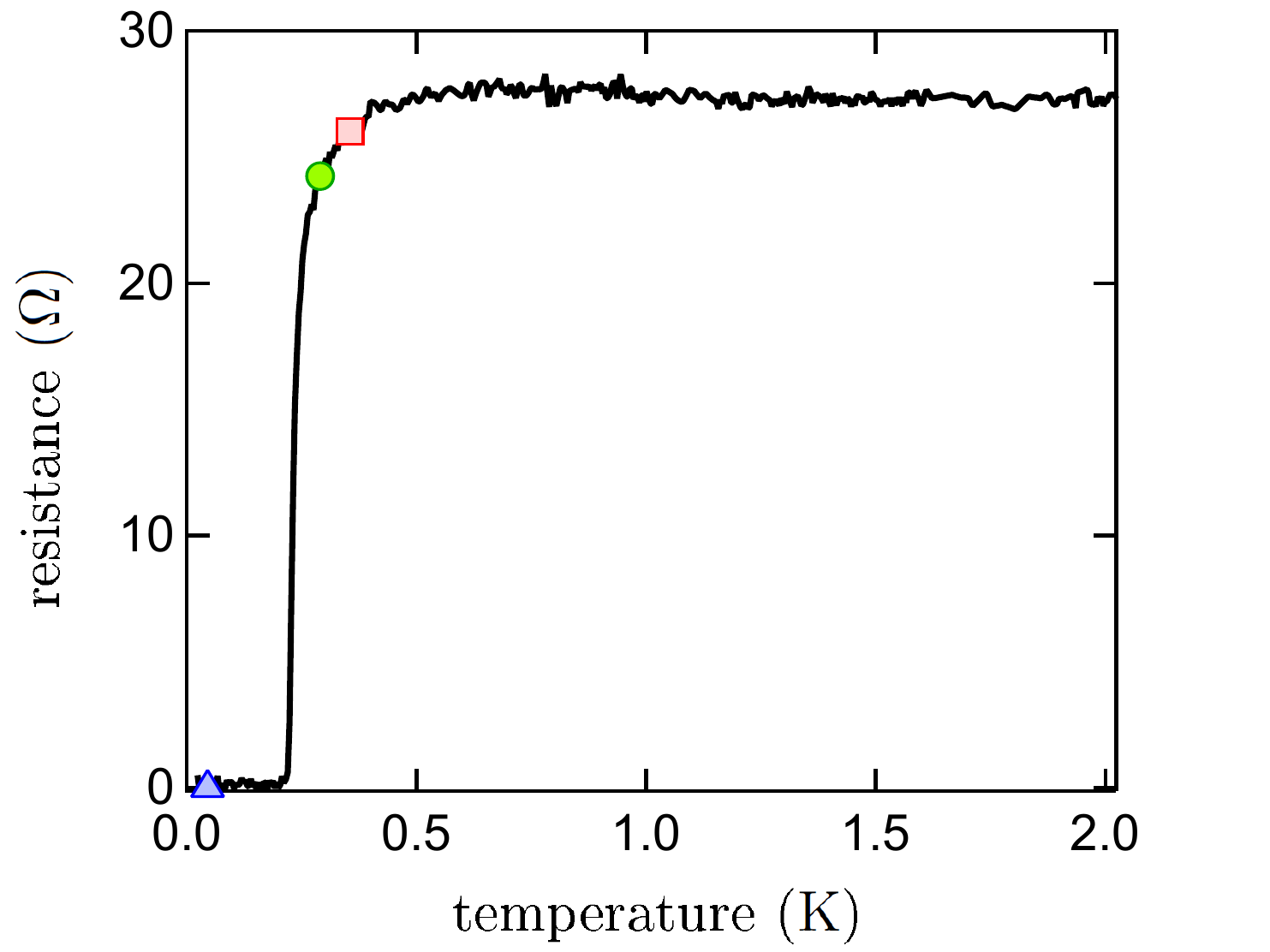}\\
  \caption{(color online) Low frequency (77 Hz) resistance as a function of temperature as measured by lock-in techniques for the a-NbSi film. The square, the circle and the triangle define the three calibration points of the SBC.}
  \label{fig:R(T)_DC}
\end{figure}

%-----------------------------------------------------------------------------------
Remarking that the error coefficients ($\alpha$, $\beta$ and $\gamma$) can be determined from any given set of three references, we propose a calibration procedure using the sample itself as reference. For this, we choose three temperatures, $T_1$, $T_2$ and $T_3$, each of them lower than $4 \, \mathrm{K}$, for which we know the sample impedance from a separate experiment or a model (see Fig.~\ref{fig:R(T)_DC}). Since the microwave response of the setup is temperature-independent at low temperature, this calibration is valid for all temperatures below $\sim$ 4 K. It is remarkable that this method also fully takes into account the sample environment, such as the lines impedance mismatch due to a varying resistance and dielectric constant at low temperature, the connectors tightening conditions, the coupler's imperfections or even the Fabry-Perot resonances due to the sample geometry (see Fig.~\ref{fig:Sample}(b)). The choice of the reference temperatures is of importance for the measurement precision. In particular, as we will see, the corresponding sample impedances have to be sufficiently different.

In the present case, the first reference point was chosen so that the sample is in the normal region. Significant superconducting fluctuations could then be neglected (see Supplementary Material). We have also checked that the reactive part of the conductance due to weak localisation is negligible within the frequency range of the measurement (see Supplementary Material). For $T_1 = 390\, \mathrm{mK}$ ($\sim 1.7 \, T_c$), we therefore assumed the sample to be purely resistive, with an impedance given by the simultaneous lock-in measurement: $Z(T_1)=Z_1=R_1=26.0 \pm 0.05\,\Omega$.

The second reference point was also chosen for the sample to be in the normal state, albeit closer to $T_c$. Given the normal state resistance value of the a-NbSi sample, we have established by varying the reference temperature $T_2$ that a minimum difference of $\Delta R=R_2-R_1\simeq2\;\Omega$ is necessary to optimize the calibration signal--to--noise ratio. For $\Delta R>2\;\Omega$, the result after calibration is unchanged, within experimental uncertainty, as will be discussed in section \ref{subsec:Error}. We therefore chose $T_2 \simeq T_1-0.1 \, \mathrm{K} = 290 \, \mathrm{mK}$. Should the normal state resistance be further away from 50 $\Omega$, a larger $\Delta R$ should be chosen. For this temperature, we have also assumed that $Z(T_2)=Z_2=R_2=24.0 \pm 0.2 \,\Omega$, given by the lock-in measurement. As the superconducting transition is approached, Aslamazov-Larkin corrections of the DC conductivity have been shown to be relevant for our system \cite{crauste2010}. However, from estimates of Aslamazov-Larkin conductivity corrections for $T>T_c$ \cite{Aslamasov1968, Ohashi2006}, we have checked that the frequency dependence of $Z(T_2)$ could be neglected in the considered frequency range (see Supplementary Material).

The third reference point was taken such that $T_3$= 25 mK $\ll T_c$. At these temperatures, usual superconductors can be assimilated to short circuits. This is the assumption made by Booth et al.\cite{Booth1994}. However, in the case of disordered superconductors such as a-NbSi, the zero-temperature kinetic inductance is large (see eq. (\ref{eq:Ls(T=0)})), so that the sample is best modeled by a pure inductance $Z(T_3)=Z_3=jL_K(T=0)\omega$. The zero-temperature kinetic inductance has been estimated using eq.~(\ref{eq:Ls(T=0)}) (see Supplementary Material). The error caused by the choice of $L_K(T=0)$ will be discussed in section \ref{subsec:Results}.
%Moreover, we have checked that a reasonable variation (\textcolor{red}{de combien?}) in the value of $L_s(T=0)$ does not qualitatively modify the obtained results.

Although the sample complex impedance is, by definition of this method, fixed at $T_1$, $T_2$ and $T_3$, its temperature and frequency dependencies close to the superconducting transition are not imposed by the calibration, as shown in the following. We can therefore infer the sample electrodynamic response at all other temperatures, provided the knowledge of the sample impedance at the three calibration points.
%-----------------------------------------------------------------------------------

%-----------------------------------------------------------------------------------
\subsection{\label{subsec:Def_planes}Definition of the calibration planes}
%-----------------------------------------------------------------------------------
As mentioned in section \ref{sec:Sample}, the sample is inserted within a transmission line resulting in Fabry-Perot type interferences. One of the main strengths of the SBC procedure is that, although it may seem counter-intuitive, it redefines the calibration planes by taking into account the Fabry-Perot resonator (see Fig. \ref{fig:Sample}(b)). Indeed, the references $Z_1$, $Z_2$ and $Z_3$ now correspond to measured reflection coefficients $\Gamma^{m}_1$, $\Gamma^{m}_2$ and $\Gamma^{m}_3$ such that:
\begin{equation}
    \Gamma^m_i=\beta+\frac{\alpha\Gamma_i}{1-\gamma \Gamma_i}\; ; \; i=\{1,2,3\}
\end{equation}
\noindent where the $\{\Gamma_i\}_{i=1,2,3}$ are defined by:
\begin{equation}
    \Gamma_i=\frac{Z_i-Z_0}{Z_i+Z_0}\; ; \; i=\{1,2,3\}
    \label{eq:Gamma_i}
\end{equation}
\noindent In other words, all the sample's environment is taken into account in the three complex coefficients $\alpha$, $\beta$ and $\gamma$. Note that the detection setup will be most efficient when $|\beta|,|\gamma| \ll 1$. By inserting these references into the calibration equations \cite{Kitano2008}, one obtains the calibration coefficients:
\begin{widetext}
\begin{eqnarray}
\label{eq:cal_coeff1}
\alpha&=&2Z_0 \frac{\Delta \Gamma_{12}^{m}\Delta \Gamma_{23}^{m}\Delta \Gamma_{31}^{m} \Delta Z_{12}\Delta Z_{23}\Delta Z_{31}}{\left(\Gamma_1^{m}\Delta Z_{10}\Delta Z_{23}+\Gamma_2^{m}\Delta Z_{20}\Delta Z_{31}+\Gamma_3^{m}\Delta Z_{30}\Delta Z_{12}\right)^2}\\
\label{eq:cal_coeff2}
\beta&=&-\frac{\Gamma_{1}^{m}\Gamma_{2}^{m}\Delta Z_{30}\Delta Z_{12}+\Gamma_{2}^{m}\Gamma_{3}^{m}\Delta Z_{10}\Delta Z_{23}+\Gamma_{3}^{m}\Gamma_{1}^{m}\Delta Z_{20}\Delta Z_{31}}{\Gamma_1^{m}\Delta Z_{10}\Delta Z_{23}+\Gamma_2^{m}\Delta Z_{20}\Delta Z_{31}+\Gamma_3^{m}\Delta Z_{30}\Delta Z_{12}}\\
\gamma&=&-\frac{\Gamma_{1}^{m}\Sigma Z_{10}^{m}\Delta Z_{23}+\Gamma_{2}^{m}\Sigma Z_{20}^{m}\Delta Z_{31}+\Gamma_{3}^{m}\Sigma Z_{30}^{m}\Delta Z_{12}}{\Gamma_1^{m}\Delta Z_{10}\Delta Z_{23}+\Gamma_2^{m}\Delta Z_{20}\Delta Z_{31}+\Gamma_3^{m}\Delta Z_{30}\Delta Z_{12}}
\label{eq:cal_coeff3}
\end{eqnarray}
\noindent where $\Delta \Gamma_{ij}^m=\Gamma_i^m-\Gamma_j^m$, $\Delta Z_{ij}=Z_i-Z_j$, $\Sigma Z_{ij}=Z_i+Z_j$ and $\{i,j\}=\{0,\,1,\,2,\,3\}$. By inverting equations (\ref{eq:Reflection_coeff}) and (\ref{eq:Gamma_sample}), we finally get the calibrated value of the impedance of the sample:
\begin{equation}
    Z(\Gamma ^m,\alpha,\beta,\gamma)=\frac{\alpha+(\Gamma^m-\beta)(\gamma+1)}{\alpha+(\Gamma^m-\beta)(\gamma-1)}
    \label{eq:Z_cal1}
\end{equation}
\noindent which can be expressed as a function of the measured values $\Gamma^m_1$, $\Gamma^m_2$, $\Gamma^m_3$ and the corresponding impedances $Z_1$, $Z_2$, $Z_3$:
\begin{equation}
    Z=  \frac{Z_1Z_2 \Delta \Gamma^m_{12}\Delta \Gamma^m_{3\star}+Z_2Z_3 \Delta \Gamma^m_{23}\Delta \Gamma^m_{1\star}+Z_3Z_1 \Delta \Gamma^m_{31}\Delta \Gamma^m_{2\star}}{Z_1 \Delta \Gamma^m_{23}\Delta \Gamma^m_{\star 1}+Z_2 \Delta \Gamma^m_{31}\Delta \Gamma^m_{\star 2}+Z_3 \Delta \Gamma^m_{12}\Delta \Gamma^m_{\star 3}}
\label{eq:Z_cal2}
\end{equation}
\end{widetext}
\noindent where $\Delta \Gamma^m_{i\star}=\Gamma^m_i-\Gamma^m$.

It is worth noting that the calibration coefficients deduced from the SC and the SBC procedures are different, as can be observed in Fig.~\ref{fig:coeff_cal}. As can be expected, the magnitude of $\beta$ is almost the same for both calibration methods since it mainly reflects the properties of the coupler. However, both $\alpha$ and $\gamma$ strongly depend on the imperfections of the transmission line on port \#3. In particular, both coefficients include the effect induced by the Fabry-Perot resonator and are therefore significantly different for $f > 500 \, \mathrm{MHz}$. We also observe standing wave patterns on the frequency dependence of $\alpha$, $\beta$ and $\gamma$. These patterns are characterized by a frequency $\delta f \sim 50 \, \mathrm{MHz}$ and correspond to multiple reflections in a $2 \, \mathrm{m}$-long cable which is consistent with the length of the detection line. These oscillations are again more pronounced in the case of the SC procedure because the calibration is performed in 4 successive coolings. In other words, the strength of the SBC procedure resides in the fact that the calibration planes flank the sample and are defined once for all measurements.

\begin{figure}
\includegraphics[width=0.85\columnwidth]{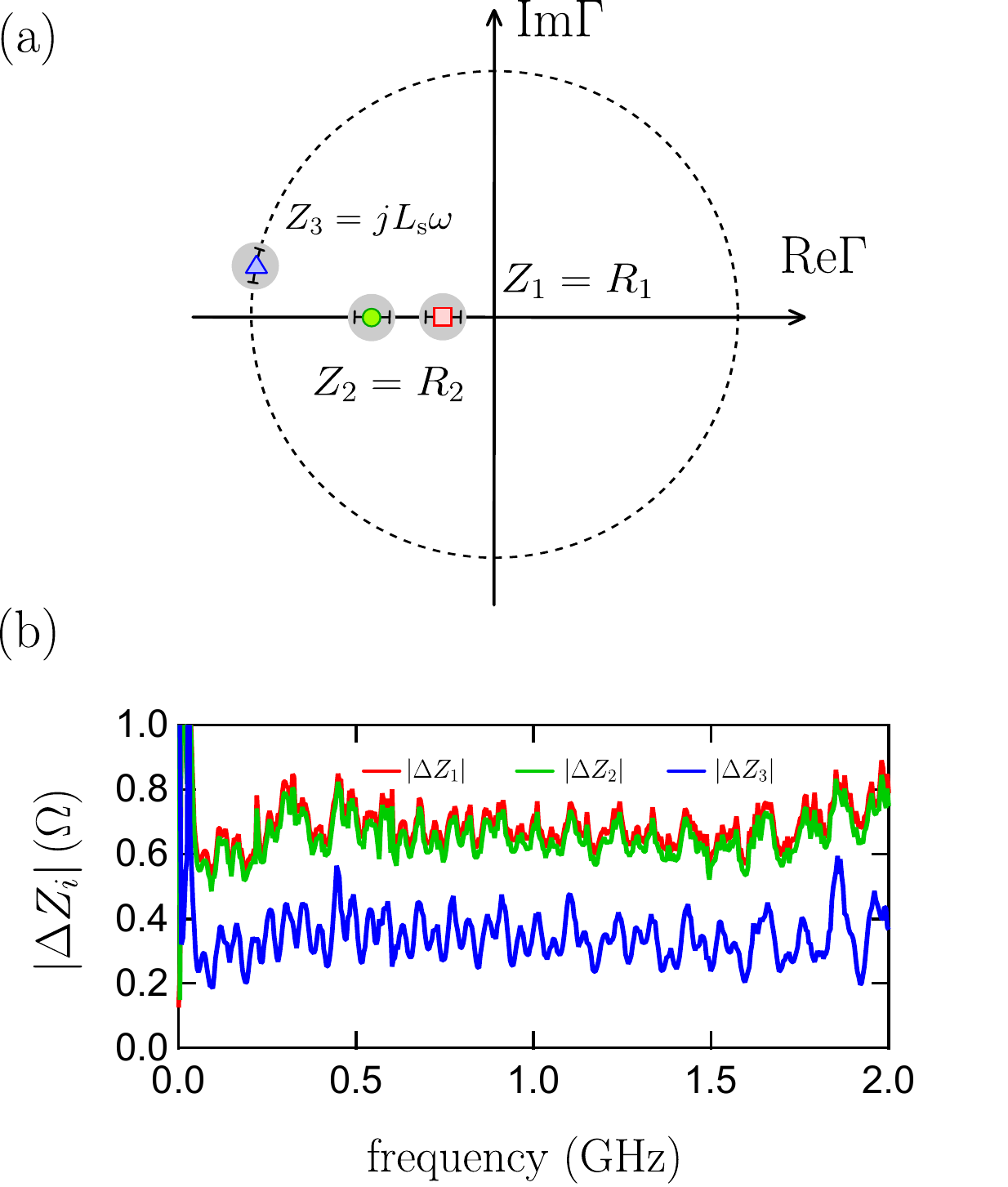}\\
  \caption{(color online) (a) Reflection coefficients $\Gamma_i$'s in the complex plane for the references $Z_1=R_1$, $Z_2=R_2$ and $Z_3=jL_K(T=0)\omega$. The SBC induces a systematic error on the calibration coefficients $\alpha$, $\beta$ and $\gamma$. This error is due to a systematic error on the $Z_i$'s (black lines) and a random error on the $\Gamma_i$'s (gray circles) due to the measurement precision. The mean values of the reflections coefficients have been shifted and the related error bars have been enlarged for the sake of clarity. (b) Minimal value of $|\Delta Z_{i}|$ for references $Z_1$, $Z_2$ and $Z_3$.}
  \label{fig:Gamma}
\end{figure}

According to equations (\ref{eq:cal_coeff1})-(\ref{eq:cal_coeff3}), the calibration coefficients are defined as soon as the references $Z_1$, $Z_2$ and $Z_3$ are different (Fig.~\ref{fig:Gamma}(a)). However, the references have to be chosen so that the related reflection coefficients $\Gamma^m_1$, $\Gamma^m_2$ and $\Gamma^m_3$ are clearly different from each others. From an experimental point of view, these coefficients are measured within a statistical uncertainty $\delta \Gamma^m$ which depends on the noise $T_N \sim 100 \, \mathrm{K}$ of the cryogenic amplifier, its gain $A \sim 43 \, \mathrm{dB}$, the VNA output power $P_{rf}$ and its resolution bandwidth $\Delta f \sim 100 \, \mathrm{Hz}$:
\begin{equation}
\delta \Gamma^m=\sqrt{\frac{Ak_BT_N\Delta f}{P_{rf}}}
\label{eq:delta_Gamma}
\end{equation}
\noindent The output power $P_{rf}$ is chosen to avoid Joule heating and its upper limit can be related to the upper limit of the DC current $I_{max} \sim 1 \, \mu \mathrm{A}$. By comparing it to the microwave power absorbed by the sample, we find:
\begin{equation}
P_{rf}\leq\frac{\mathrm{Re}(Z)I_{max}^2}{|\alpha_e|^2(1-|\Gamma(Z)|^2)} \leq 200 \, \mathrm{nW}
\end{equation}
\noindent where  $|\alpha_e|^2=-40 \, \mathrm{dB}$ stands for the attenuation on the excitation line and $Z$ has been evaluated in the normal state ($T=$ 4 K). In the experiment, the excitation will be set to $P_{rf}= 100 \, \mathrm{nW}$, \textit{i.e.} $10 \, \mathrm{fW}$ on the sample. This gives rise to a statistical uncertainty $\delta \Gamma^m \sim 5 \times 10^{-3}$. The determination of the calibration coefficients thus requires a minimal difference $|\Delta Z_{min}| \gg |\Delta Z_{i}|= |dZ/d\Gamma^m| \delta \Gamma_m$ between the impedances of references $Z_1$, $Z_2$ and $Z_3$:
\begin{equation}
|\Delta Z_{i}| \simeq \frac{\left|\Sigma Z_{i0}-\gamma \Delta Z_{i0}\right|^2}{2|\alpha| Z_0}  \delta \Gamma_m \; ; \; i=\{1,2,3\}
\label{eq:delta_Zi}
\end{equation}
\noindent Figure~\ref{fig:Gamma}(b) shows $|\Delta Z_{i}|_{i=1,2,3}$ as a function of frequency using the calibration coefficients deduced from the SBC. The various $|\Delta Z_{i}|$ are essentially frequency-independent, except for the oscillating patterns, of caracteristic frequency 50 MHz, which are due to the standing waves along the 2 m-long cable of the detection line. From Eq.~\ref{eq:delta_Zi}, the frequency dependence of $|\Delta Z_{i}|$ is indeed not only due to the coefficient $\alpha$ but also to $\gamma$. It is reasonable to assume that for all standards, $|\Delta Z_{i}| < 800 \, \mathrm{m}\Omega$. For security, we define a criterion for the choice of the references: they must be such that $|\Delta Z_{min}| \sim 1 \, \Omega$. The references described in section \ref{subsec:Gal_pple} were such that $\Delta R >2 \, \Omega$, which conforms to the criterion.

%-----------------------------------------------------------------------------------

%-----------------------------------------------------------------------------------
\subsection{\label{subsec:Error}Error on the calibration}
%-----------------------------------------------------------------------------------

As previously discussed, a minimal difference $|\Delta Z_{min}|$ between standards is necessary to perform the calibration because of the random errors in the experimental measurements. However, a large difference in the references impedances $\Delta Z_{ij}$ does not mean the errors are small. The errors $\delta \alpha$, $\delta \beta$ and $\delta \gamma$ on calibration coefficients indeed induce \textit{systematic errors} on impedance measurements. This errors read:
\begin{equation}
\delta \kappa = \sum_{i=1}^{3} \left(\frac{\partial \kappa}{\partial \Gamma^m_i} \right) \delta \Gamma^m_i + \sum_{i=1}^{3} \left(\frac{\partial \kappa}{\partial Z_i} \right) \delta Z_i \; ; \; \kappa=\{\alpha,\beta,\gamma\}
\label{eq:err_coeff_cal}
\end{equation}
\noindent where $\delta \Gamma^m_i$ is the random error on the reflection coefficients $\Gamma^m_i$ and $\delta Z_i$ is the systematic error on the reference impedances $Z_i$. To minimize $\delta \Gamma^m_i$, several measurements of the standards have been averaged as the temperature was slowing ramping (see Suplementary Material), so that it in turns introduces a small contribution to $\delta Z_i$ (for references $Z_1$ and $Z_2$). $\delta \kappa$ therefore includes the statistical uncertainty in the measurement of the standard response, the error introduced by a non-stabilized temperature and the error due to the assumptions made on $Z_i$.

$\delta \alpha$, $\delta \beta$ and $\delta \gamma$ are complex numbers, they are frequency--dependent and change for each calibration procedure. In the following, the calibration coefficients are averaged over 10 measurements (see Supplementary Material) to improve the quality of the calibration giving rise to:
\begin{eqnarray}
|\delta \Gamma^m_1| &\sim&  |\delta \Gamma^m_2|  \sim  |\delta \Gamma^m_3|  \sim   \delta \Gamma^m/\sqrt{10} \sim  5 \times 10^{-4} \label{eq:dcal1}\\
\delta Z_1  &\sim &  50 \, \mathrm{m}\Omega, \; \delta Z_2  \sim  200 \, \mathrm{m}\Omega, \;\delta Z_3  = 0 \, \Omega \label{eq:dcal2}
\end{eqnarray}
\noindent We will discuss the choice of $Z_3$ in the next section (see also Suplementary Material). Our goal here is to determine the maximum error $|\delta Z_{max}|$ that one makes on the sample impedance using a given calibration procedure. By considering the random errors on the reflectometry measurements and the systematic errors on the calibration coefficients, we deduce from the root mean square error:
\begin{widetext}
\begin{eqnarray}
|\delta Z_{max}|^2= \left| \frac{\partial Z}{\partial \Gamma^m} \right|^2 \left( \delta \Gamma^m \right)^2 + \sum_{i=1}^{3} \left| \sum_{\kappa=\alpha,\beta,\gamma} \frac{\partial Z}{\partial  \kappa} \frac{\partial \kappa}{\partial  \Gamma^m_i} \right|^2 \left(\frac{\delta \Gamma^m}{\sqrt{10}}\right)^2 + \sum_{i=1}^{3} \left| \sum_{\kappa=\alpha,\beta,\gamma}\frac{\partial Z}{\partial  \kappa} \frac{\partial \kappa}{\partial  Z_i} \right|^2 \left|\delta Z_i\right|^2
\label{eq:dZ}
\end{eqnarray}
\noindent The first term corresponds to the statistical reproducibility of the measurement. The other terms are related to the systematic errors coming from the determination of the calibration coefficients. This quantity can be expressed as a function of the calibration coefficients and the measured impedance:

\begin{align}
|\delta Z_{max}|^2=  \left| \frac{\left(\Sigma Z_{i0}-\gamma \Delta Z_{i0}\right)^2}{2|\alpha| Z_0}  \right|^2 \left( \delta \Gamma^m \right)^2 + \sum_{(i,j,k)} \left| \frac{\left(\Sigma Z_{i0}-\gamma \Delta Z_{i0}\right)^2}{2|\alpha| Z_0} \frac{\Delta Z_{\star j}\Delta Z_{\star k}}{\Delta Z_{ij}\Delta Z_{ik}}  \right|^2 \left(\frac{\delta \Gamma^m}{\sqrt{10}}\right)^2 + \sum_{(i,i,k)} \left| \frac{\Delta Z_{\star j}\Delta Z_{\star k}}{\Delta Z_{ij}\Delta Z_{ik}}  \right|^2 \left|\delta Z_i\right|^2
\label{eq:dZ2}
\end{align}

%\begin{eqnarray}
%|\delta Z_{max}|^2=  \left| \frac{\left(\Sigma Z_{i0}-\gamma \Delta Z_{i0}\right)^2}{2|\alpha| Z_0}  \right|^2 \left( \delta \Gamma^m \right)^2 + \sum_{(i,j,k)} \left| \frac{\left(\Sigma Z_{i0}-\gamma \Delta Z_{i0}\right)^2}{2|\alpha| Z_0} \frac{\Delta Z_{\star j}\Delta Z_{\star k}}{\Delta Z_{ij}\Delta Z_{ik}}  \right|^2 \left(\frac{\delta \Gamma^m}{\sqrt{10}}\right)^2 + \sum_{(i,i,k)} \left| \frac{\Delta Z_{\star j}\Delta Z_{\star k}}{\Delta Z_{ij}\Delta Z_{ik}}  \right|^2 \left|\delta Z_i\right|^2
%\label{eq:dZ2}
%\end{eqnarray}
\end{widetext}
\noindent where $\Delta Z_{i\star}=Z_i-Z$ and $(i,j,k)=\{(1,2,3);(2,3,1);(3,1,2)\}$.

\begin{figure}
\includegraphics[width=0.9\columnwidth]{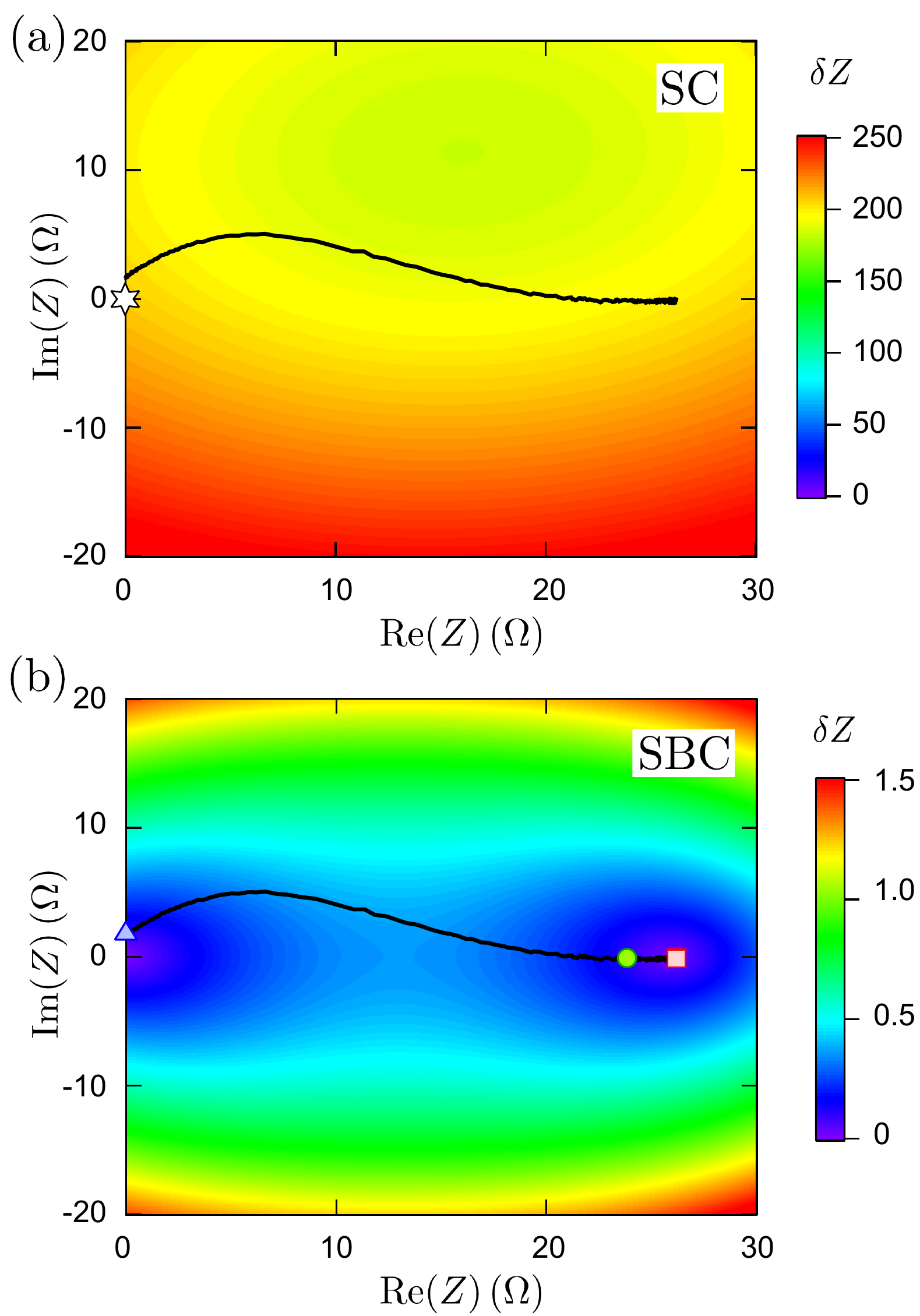}\\
  \caption{(color online) Color map of the uncertainty $|\delta Z|$ on the impedance for (a) the SC and (b) the SBC procedures. The solid lines corresponds to the evolution of the sample impedance $Z(2 \, \mathrm{GHz})$ as temperature goes from $20 \, \mathrm{mK}$ to $400 \, \mathrm{mK}$. For the SBC, the square, the circle and the triangle define the three calibration points: $Z_1=26.0\, \Omega$, $Z_2=24.0\, \Omega$ and $Z_3=jL_K(T=0) \omega$ respectively. For the SC, the star defines the ``short'' calibration point. The ``open'' and the ``load'' points are not visible in the considered impedance range.}
  \label{fig:err}
\end{figure}

A numerical calculation using the calibration coefficients derived from both the SC and the SBC procedures enables to return the error $|\delta Z_{max}|$ as a color map in the impedance range of interest (Fig.~\ref{fig:err}). In the SC procedure, the three standards and the sample have been cooled down successively. The random error on the reflection coefficient is then no longer defined by the amplifier noise, but by the reproducibility of the cool-downs. Because we are working at dilution refrigerator temperatures, as mentioned in section~\ref{sec:Setup}, numerous microwave components need to be inserted along the measurement setup (see Supplementary Material). The reflection of the line can then unavoidably differ by $3\, \mathrm{dB}$ from one cooling to another (see Supplementary Material). When working at millikelvin temperature, it is thus preferable to develop a single-cooldown calibration. The random error on the reflection coefficient is thus  $\delta \Gamma^m \sim \sqrt{2}$. Then, the error on the impedance measurement (Fig.~\ref{fig:err}(a)) can reach $|\delta Z_{max}| \sim 200 \, \Omega$ as observed in Fig.~\ref{fig:Z_SC}. Figure~ \ref{fig:err}(b) shows $|\delta Z_{max}|$ for the SBC procedure. It has been calculated by using the random errors on the refection coefficients and the systematic errors on the references as previously determined (see equations (\ref{eq:dcal1}), (\ref{eq:dcal2}) and (\ref{eq:dZ2})). Although the error on the SC could probably be minimized by optimizing the experimental set-up for this specific calibration, there is almost a difference of three orders of magnitude in the obtained $|\delta Z_{max}|$ between the two calibrations. For the SBC, the upper bound of the error $|\delta Z_{max}| \leq 0.5 \, \Omega$ will allow us to compare the calibrated data with theoretical expectations. The error $|\delta Z_{max}|$ in the SBC procedure is, as expected, minimal around the references impedances. This is not observed in the SC procedure because the calibration coefficients are deduced from three different coolings and do not exactly describe the microwave setup used to measure the sample. Note that if it had been possible to chose well separated reference impedances in the SBC procedure (as ``open'', ``short'' and ``load'') the precision of the measurement would have been better by one order of magnitude $|\delta Z_{max}| \leq 0.05 \, \Omega$.

%-----------------------------------------------------------------------------------

%-----------------------------------------------------------------------------------
\subsection{\label{subsec:Results}Results}
%-----------------------------------------------------------------------------------

\begin{figure}[!h]
\includegraphics[width=0.9\columnwidth]{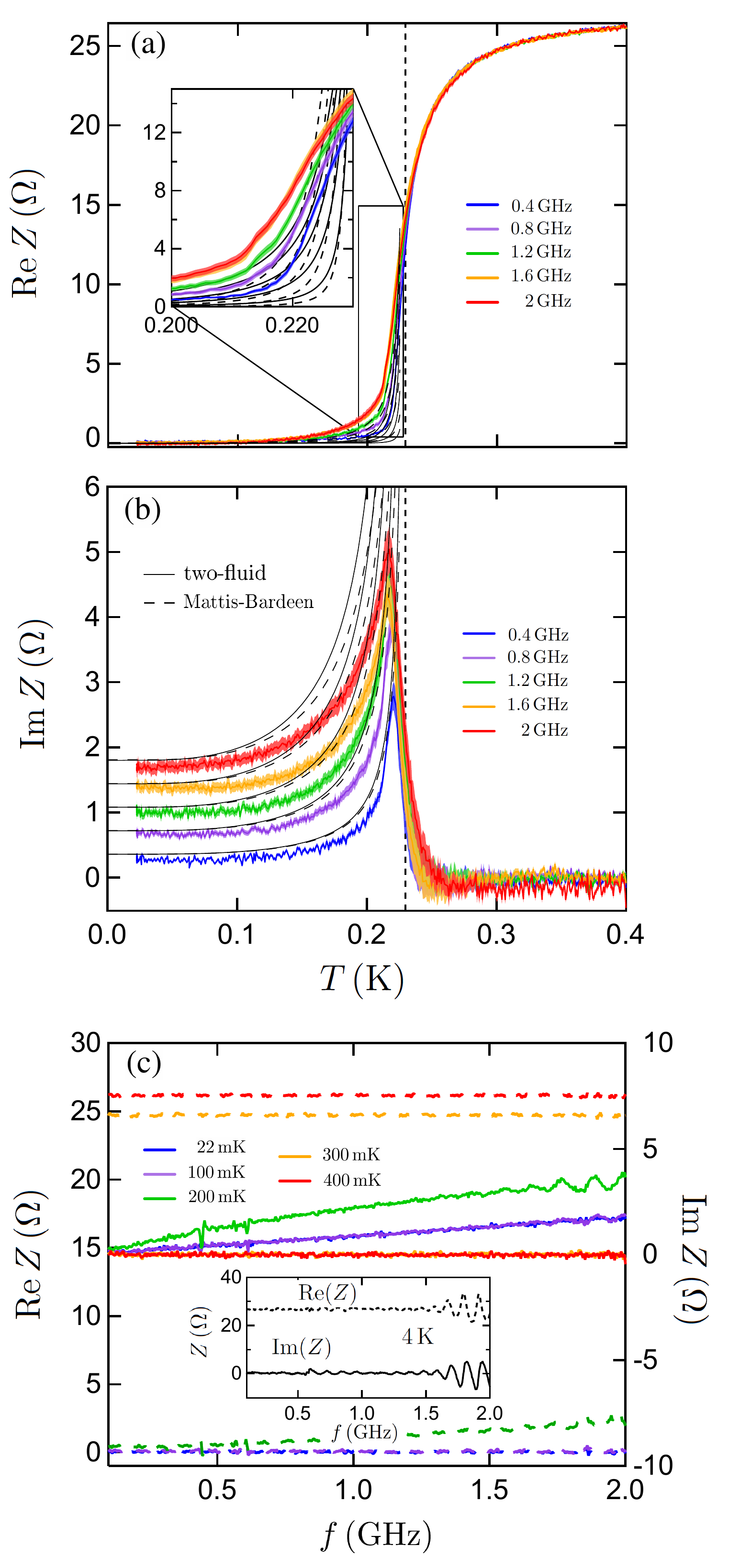}\\
  \caption{(color online) (a) Real and (b) imaginary parts of the complex impedance (color dots) as a function of temperature for selected frequencies, as obtained from the SBC (see text). For temperatures lower than $T_c$, the solid black lines correspond to $\mathrm{Re}(Z)$ and $\mathrm{Im}(Z)$ as given by equations (\ref{eq:RL1}), (\ref{eq:RL2}) and (\ref{eq:Z_BCS}). The dashed lines correspond to Mattis-Bardeen theory. \textit{Inset of figure (a):} zoomed-up plot of the data and theoretical expectations. (c) Real (dashed lines) and imaginary (solid lines) parts of the complex impedance as a function of frequency for selected temperatures. Inset : Complex impedance obtained using the same calibration on data taken at 4 K thirteen day prior to the calibration.}
  \label{fig:Z_SBC}
\end{figure}

The sample impedance obtained after SBC is shown as a function of temperature and frequency in Fig.~\ref{fig:Z_SBC}. Before analyzing the evolution of the sample impedance in temperature, let us point out the experimental robustness of the SBC method: Fig.~\ref{fig:Z_SBC}.c shows the sample impedance as a function of frequency, at five selected temperatures. The inset of the same figure gives the sample impedance computed using the same calibration on data that were obtained at 4 K \textit{thirteen days} before. As can be seen, the 4 K impedance is consistent with the DC measurement and the estimated errors except at frequencies higher than 1.5 GHz, probably due to the overall evolution of the measurement set-up.
The shaded areas in Fig.~\ref{fig:Z_SBC}.a and b   correspond to the error $|\delta Z_{max}|$ . As expected, in the normal state, the impedance is purely real. At the superconducting transition, the maximum amplitude of Im($Z$) increases with frequency, due to the finite superfluid density developing at $T<T_c$. Let us emphasize that the calibration imposes the sample response in the normal state at $T_1, T_2 > T_c$ and in the superconducting state at $T_3\ll T_c$. The divergence of the kinetic inductance responsible for the Im($Z$) peak thus cannot be explained by a calibration artefact.

%\begin{figure}
%\includegraphics[width=0.9\columnwidth]{Fig_Z_SBC}\\
%  \caption{(color online) (a) Real and (b) imaginary parts of the complex impedance (color dots) as a function of temperature for selected frequencies, as obtained from the SBC (see text). For temperatures lower than $T_c$, the solid black lines correspond to $\mathrm{Re}(Z)$ and $\mathrm{Im}(Z)$ as given by equations (\ref{eq:RL1}), (\ref{eq:RL2}) and (\ref{eq:Z_BCS}). The dashed lines correspond to Mattis-Bardeen theory. \textit{Inset of figure (a):} zoomed-up plot of the data and theoretical expectations.}
%  \label{fig:Z_SBC}
%\end{figure}

\begin{figure}
\includegraphics[width=0.9\columnwidth]{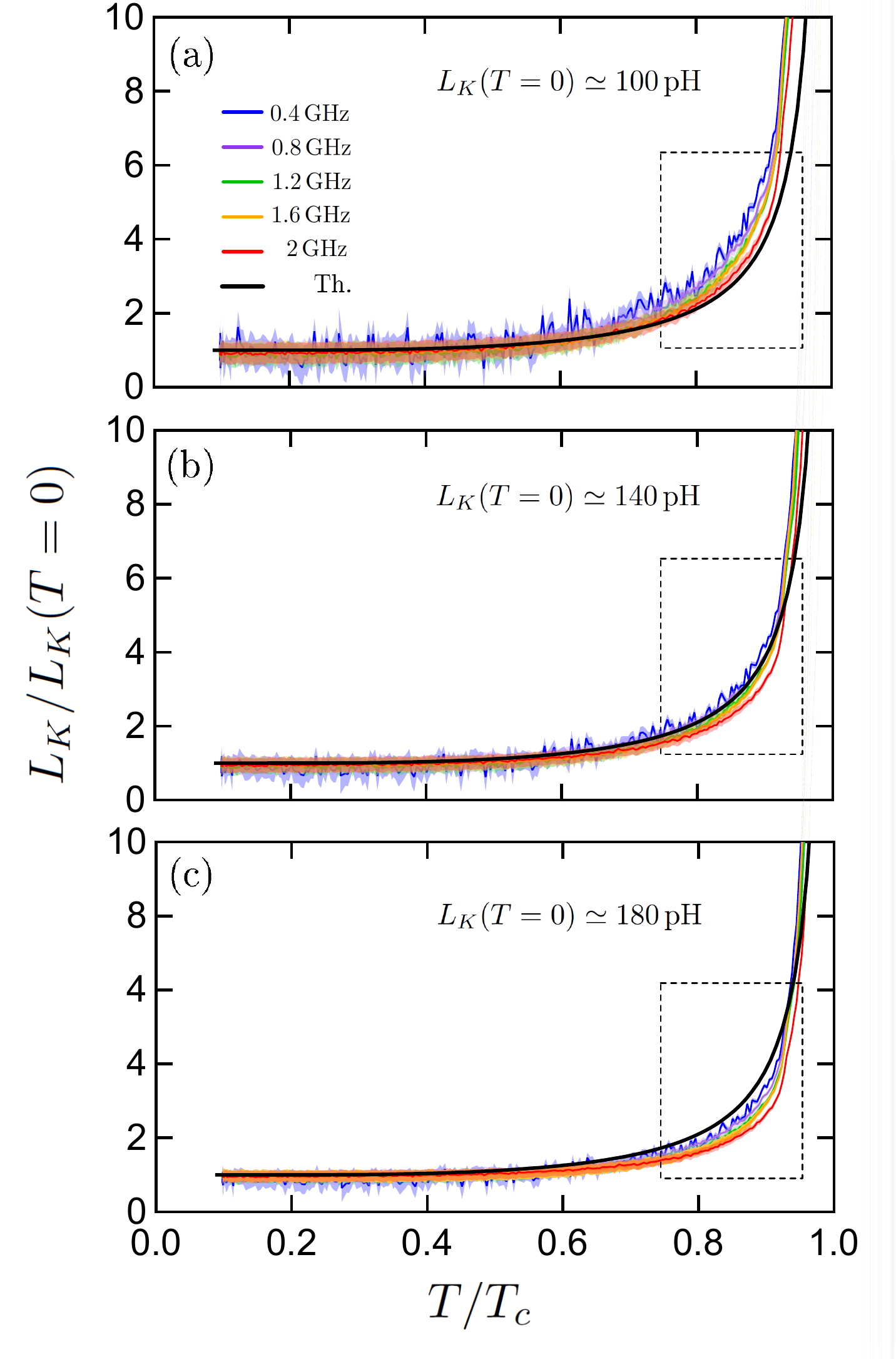}\\
  \caption{(color online) Kinetic inductance deduced from the measurement of $\mathrm{Im}(Z)$ for selected frequencies (color dots) for different values of the reference $L_K(T=0)$ : $100 \, \mathrm{pH}$ (a), $140 \, \mathrm{pH}$ (b) and $180 \, \mathrm{pH}$ (c). The solid line correspond to the theoretical expectation in the low frequency limit given by equation (\ref{eq:RL2}).}
  \label{fig:Lk}
\end{figure}

In Fig.~\ref{fig:Lk}, the kinetic inductance is plotted as a function of temperature, assuming that the reactive part of the electromagnetic response of the superconductor is well described by an inductance (see equation (\ref{eq:RL2})): $L_K=\left(\omega \mathrm{Im}(1/Z)\right)^{-1}$. This quantity clearly depends on the choice of the calibration reference $Z_3=jL_K(T=0)\omega$. In order to assess the systematic error made when assigning a given value to $L_K(T=0)$, we have plotted in the different panels the result of the calibration for $L_K(T=0) \simeq 100 \, \mathrm{pH}$, $140 \, \mathrm{pH}$ and $180 \, \mathrm{pH}$. These values correspond to a superconducting gap of $35 \, \mu\mathrm{eV} \, \pm \, 30\%$, as has been measured on the same system by Scanning Tunneling Microscopy\footnote{C. Chapelier, C. Tonnoir, E. Driessen, \textit{unpublished}.}. When comparing the results of the calibration (color dots) with BCS theory (solid black line) on the whole temperature range, it is clear that the temperature dependence of $L_K(T)$ is unaffected by the choice of $L_K(T=0)$, as is natural due to the influence of the reference $Z_3$. However, for $T \gtrsim 0.9 T_c$, there is a deviation from the theory for $L_K(T=0) \simeq 100 \, \mathrm{pH}$ and $180 \, \mathrm{pH}$ larger than the estimated maximum error (dashed rectangles in Fig.~\ref{fig:Lk}). The temperature evolution of $L_K$ hence allows us to justify \textit{a posteriori} the choice of the calibration reference $Z_3=j\omega \hbar R_{n}/\pi\Delta_0 \simeq 140 \, \mathrm{pH}$ given by the BCS theory with an average gap value. In the following, we will fix $L_K(T=0)=140 \, \mathrm{pH}$ for fixing the impedance value $Z_3$.

Let us now compare the results extracted from the SBC with the theoretical predictions for the temperature and frequency dependences of the sample impedance. The solid black lines in Fig.~\ref{fig:Z_SBC} corresponds to the theoretical expectation given by the two-fluid model where the BCS temperature dependence of the gap has been introduced (equations (\ref{eq:RL1}), (\ref{eq:RL2}) and (\ref{eq:Z_BCS})). While the qualitative features of the data can be reasonably well explained by this simple model, in particular at low frequency and far from $T_c$, there is a discrepancy close to the sample critical temperature, both for $\mathrm{Re}(Z)$ and $\mathrm{Im}(Z)$. The increase of $\mathrm{Re}(Z)$ with the frequency cannot be accounted for by a sole inductance. As shown in Fig.~\ref{fig:Z_SBC}(a), the two-fluid model resistance $R(T)$ deduced from the normal state (solid line) does not describe the experimental data and the dissipation related to thermally excited quasiparticles has to be taken into account. This dissipation corresponds to the coherence peak appearing close the $T_c$ in the real part of the Mattis-Bardeen conductance and could explain that we underestimate $\mathrm{Re}(Z)$. However, the comparison with the Mattis-Bardeen theory\cite{Mattis1958} (black dashed lines) -- which should take the quasiparticles dissipation into account -- does not yield a better agreement.

Thanks to the high precision on our measurements (see section \ref{subsec:Error}), we can confidently say that the complex impedance of the superconducting a-NbSi thin film therefore departs from the theoretical expectations of BCS theory close to $T_c$. There can be two main reasons for this. The first is an error in the assumptions we have made for $Z_1$, $Z_2$ and $Z_3$. As has been detailed previously, we have checked that the assumptions that $Z_1$ and $Z_2$ were pure resistances whereas $Z_3$ could be assimilated to a pure inductance were justified (see also Supplementary Material). Moreover, the electromagnetic model for the sample (Fig.~\ref{fig:Sample}(b)) has been tested and we have checked in a separate experiment that the Pt buffer layer in between the gold transmission line and the sample did not give rise to unaccounted dissipation effects. However, since these experiments are very sensitive, we cannot exclude spurious effects. The theory/experiment discrepancy could however also be the consequence of the disordered nature of these films. Indeed, thin a-NbSi films have been shown to exhibit disorder-induced Superconductor--to--Metal--to--Insulator Quantum Phase Transitions (QPT)\cite{Crauste2013, Couedo2016}. The theory of finite size scaling predicts a generalized scaling relation for the conductivity and thus an additional $h f/k_BT$ dependence in the conductance close to a QPT\cite{Sondhi1997}. This dependence has been observed experimentally in a-NbSi across the Metal--to--Insulator Transition\cite{Lee2000}. Although the a-NbSi sample we study here is at a certain distance from the quantum critical point of the Superconductor--Insulator Transition, the QPT could indeed still manifest itself at $T>0$. Moreover, such disordered films are known to exhibit Berezinskii--Kosterlitz--Thouless--type physics resulting in a depression of $L_K(T)$ close to $T_c$, for temperatures larger than the characteristic temperature $T_{BKT}$, due to frequency-dependent phase fluctuations\cite{Mondal2011, Liu2011}. These issues should be addressed in a systematic study of the electrodynamic response of a-NbSi superconducting thin films when disorder is varied. We would need to approach the critical points and cross to the metallic or insulating phases by studying samples with different niobium concentrations for instance. This will be the subject of a subsequent work.

%-----------------------------------------------------------------------------------
%################################################################################################

%################################################################################################
\section{\label{sec:Conlusion}Conclusion}
%################################################################################################
In order to circumvent the issue of the low temperature calibration for microwave broadband measurements, we propose a calibration method which is based on the knowledge, whether experimental or theoretical, of the sample impedance at three different values of a tunable parameter. In the present work, we made use of the temperature dependence of the resistance, but the method could be extended to exploit a known magnetoresistance or the superconducting sample critical current for instance. Using these references, one can infer the absolute value of the sample complex impedance at all other temperatures, with an enhanced sensitivity due to the fact that this calibration fully takes into account the sample electromagnetic environment, provided the set-up is temperature-independent. Moreover, it only requires a single cool-down which makes the experiments easier to perform. 

We would like to emphasize that any compound having a variation in impedance with any external parameter (not only the temperature but the electric field, the current, the magnetic field,....) could benefit from this calibration method. For instance, magnetic compounds, Josephson junction arrays \cite{van1992field} or mesoscopic circuits \cite{Gabelli2006} could be candidates. We could also think of 2D systems like graphene-based samples  \cite{han2014collapse} and superconducting interfaces \cite{Singh2018}, where a gate electric field can be used as a tuning parameter for the calibration. We therefore believe that this sample-based calibration method is very promising for characterizing the frequency dependence of condensed matter systems and could complement previously developed calibration methods.

%################################################################################################

\section*{\label{sec:Supplementary Material}Supplementary Material}

See supplementary material for details on the microwaves lines of the set-up and the data acquisition protocole. Additional information are also given on the references that we used for the Standard Calibration (SC) as well as a discussion on the random error on the reflection coefficient for this SC. The assumptions made to define the references for the Sample-Based Calibration (SBC) are also precised.

%###########################################################################################
% If in two-column mode, this environment will change to single-column format so that long equations can be displayed.
%\begin{widetext}
%$$\mbox{put long equation here}$$
%\end{widetext}

% Figures should be put into the text as floats.
% Use the graphics or graphicx packages (distributed with LaTeX2e).
% Put the label that you will use with \ref{} command in the braces of the \label{} command.
%
% \begin{figure}
% \end{figure}

% \begin{table}
% \end{tabular}
% \end{table}

\color{black}

\begin{acknowledgments}
This work has been partially supported by the ANR (grant No. ANR 2010 BLAN 0403 01) and by the Triangle de la Physique (grant No. 2009-019T-TSI2D).
\end{acknowledgments}

%§§§§§§§§§§§§§§§§§§§§§§§§§§§§§§§§§§§§§§§§§§§§§§§§§§§§§§§§§§§§§§§§§§§§§§§§§§§§§§§§§§§§§§§§§§§§§§§§§§§§§§§§§§
%Bibliography
	%modif arxiv

%\bibliography{2018_01_1_RF_Francois}
%§§§§§§§§§§§§§§§§§§§§§§§§§§§§§§§§§§§§§§§§§§§§§§§§§§§§§§§§§§§§§§§§§§§§§§§§

\end{document}